\documentclass[nocopyrightspace]{sig-alternate}
\usepackage{times,amsfonts,amssymb,amsmath,comment}
\usepackage{bbm,graphicx,url,bm,cite}
\usepackage{subfigure}
\usepackage{psfrag}

\newcommand{\enne}{{\ensuremath{\mathbb{N}}}}
\newcommand{\equaref}[1]{(\ref{eq:#1})}

\newcommand{\Os}
{
\hbox{$\Theta$}
}

\newcommand{\diff}{{\rm\,d}}

\newtheorem{lemma}{\bf Lemma}

\newtheorem{proposizione}{\bf Proposition}

\newcommand{\ls}[1]
   {\dimen0=\fontdimen6\the\font
    \lineskip=#1\dimen0
    \advance\lineskip.5\fontdimen5\the\font
    \advance\lineskip-\dimen0
    \lineskiplimit=.9\lineskip
    \baselineskip=\lineskip
    \advance\baselineskip\dimen0
    \normallineskip\lineskip
    \normallineskiplimit\lineskiplimit
    \normalbaselineskip\baselineskip
    \ignorespaces
}

\newcommand{\tgifeps}[3]{
\begin{figure}[tb]
\centering
\includegraphics[width=#1cm]{#2.eps}
\caption{#3\label{fig:#2}}
\end{figure}
}

\excludecomment{lunga}
\includecomment{corta}

\begin{document}

%\conferenceinfo{MobiCom'04,} {Sept. 26-Oct. 1, 2004, Philadelphia, Pennsylvania, USA.}
%\CopyrightYear{2004}
%\crdata{1-58113-868-7/04/0009}

\begin{sloppypar}
\bibliographystyle{plain}
\nocite{*}

\title{Restricted Mobility Improves Delay-Throughput Trade-offs
in Mobile Ad-Hoc Networks}
%\title{Capacity Scaling in Ad Hoc Wireless Networks with Heterogeneous Mobile Nodes
%\begin{picture}(10,10)(335,75)
%{\small Submitted to ACM MobiHoc 2008}
%\end{picture}

%\subtitle{Paper \#1569085411 submitted to MobiHoc 2008}
%\authorinfo{Paper \#1569109605 submitted to MobiCom 2008 \vspace{-15mm}}
%\author{Paper \#1569109605 submitted to MobiCom 2008 \vspace{-1mm}}

\author{ Michele Garetto $\!^\dagger \quad$  Emilio Leonardi $\!^\ast$  \vspace{3mm} \\
$^\dagger$ Dipartimento di Informatica, Universit\`a di Torino, Italy \vspace{1mm} \\
$^\ast$ Dipartimento di Elettronica, Politecnico di Torino, Italy }

%\thanks{$^\ast$ M. Garetto is with Dipartimento di Informatica, Universit\`a di
%Torino, Italy; P. Giaccone and E. Leonardi are with
%Dipartimento di Elettronica, Politecnico di Torino, Italy }}

\maketitle
%\ls{0.75}
\begin{abstract}
In this paper, we analyze asymptotic delay-throughput trade-offs
in mobile ad-hoc networks comprising heterogeneous nodes with restricted
mobility. We show that node spatial heterogeneity has the ability to drastically improve
upon existing scaling laws established under the assumption
that nodes are identical and uniformly visit the entire
network area. In particular, we consider the situation in which each node
moves around its own {\em home-point} according to a
restricted mobility process which results into a
spatial stationary distribution that decays
as a power law of exponent $\delta$ with the distance from the home-point.
For such restricted mobility model, we propose a novel class of
scheduling and routing schemes, which significantly outperforms
all delay-throughput results previously obtained in the case
of identical nodes. In particular, for $\delta = 2$ it is
possible to achieve almost constant delay and almost constant
per-node throughput (except for a poly-logarithmic factor)
as the number of nodes increases, even without resorting to
sophisticated coding or signal processing
techniques.
\end{abstract}
%\ls{0.85}

\section{Introduction}\label{sec:intro}

Over the last decade we have seen a flurry of theoretical studies
aimed at establishing fundamental scaling laws
of ad-hoc networks as the number of nodes increases.
Gupta and Kumar first considered the case of $n$ static nodes
and $n$ random source-destination (S-D) pairs, obtaining
the disheartening result that the maximum per-node throughput
decays at least as $1/\sqrt{n}$ \cite{Gupta-Kumar}.
\begin{lunga}
{\bf In \cite{franceschetti} it is shown that the above bound
is indeed achievable even in the case in which nodes are randomly
placed over the network area.}
\end{lunga}

In contrast to static networks, Gossglauser and Tse~\cite{Grossglauser-Tse} have
shown that a constant per-node throughput can be achieved in
mobile ad-hoc networks by exploiting the {\em store-carry-forward}
communication paradigm, i.e., by allowing nodes to store the data and physically carry
them while moving around the network.
\begin{lunga}
\bf  Of course this nice scalability property comes at the
expense of significant data transfer delays, on the time scale of nodes' movements
across the network area, which, however, can be considered acceptable in the context
of Delay Tolerant Networking \cite{dtnrg}. }
\end{lunga}
The result in \cite{Grossglauser-Tse} was proven under the
assumption that nodes independently move according to a generic, ergodic mobility
process which results, for each node, into a uniform stationary distribution
over the space. This mobility model is actually a generous one, as it allows
each node to equally come in contact with any other node, achieving
a full, homogeneous mixing.

In practical cases, however, the mobility pattern of
individual nodes is expected to be restricted over the network area, as users spend most
of the time in proximity of a few preferred places \cite{kang}, and rarely go outside
their region of habit.
%Although any two nodes are likely, in the long run,
%to eventually come in contact with each other, the impact
%of rare contacts on the overall network capacity has to be carefully
%investigated.
This observation has already motivated some researchers to study the impact
of restricted mobility models. In \cite{Grossglauser-TseII} a one-dimensional mobility model
is considered, in which each node uniformly visits a randomly chosen great circle on the unit
sphere, obtaining again a constant throughput.
%In \cite{moraes} the network of unit area is partitioned into square cells, and nodes
%are restricted to move within one randomly chosen cell; the authors consider
%two cases in which the cell area either scales as $(\log n)/n$ or remains constant,
%obtaining a throughout similar to Gupta-Kumar and Grossglauser-Tse, respectively.
In \cite{mobihoc07,infocom08} the authors consider a two-dimensional, restricted mobility model
which produces, for each node, a rotationally invariant spatial distribution
centered at a home-point uniformly chosen in the area; the resulting throughput
varies with continuity in between the two extreme cases of static nodes (Gupta-Kumar)
and fully mobile nodes (Grossglauser-Tse), depending on how the physical
network extension scales with respect to the average distance reached by the
nodes from their home-point. This result confirms that throughput is maximized when
the nodes span the entire extension of the network area. However, the authors of \cite{mobihoc07,infocom08}
have not analyzed the delay under their restricted mobility model.

Driven by the optimality of the homogeneous mixing assumption in terms of
throughput, many authors have analyzed asymptotic delay-throughput trade-offs
under the same assumption. This choice is also motivated by the fact that the most
popular mobility models adopted in the literature (such as random walk, random way-point)
produce a uniform stationary distribution over the area.\footnote{the stationary distribution of a node
under the random way-point is uniform in the absence of border effects, such as on the surface
of a sphere or a torus.} Indeed, when considering also the data transfer delay, the precise details on how
the nodes move become important. Several mobility models have been analyzed,
ranging from the simple reshuffling model \cite{Toumpis}, to the Brownian motion \cite{shah1},
and variants of random walk and random way-point \cite{bansal,sharma}.
In all of these studies, nodes have been assumed to be identical and fully mobile, i.e.,
their trajectories \lq fill the space' over time, uniformly visiting the entire
network area.

Starting from a different perspective, essentially aimed at reducing the large
delays encountered in sparse, intermittently connected mobile ad hoc networks, many researchers
have already proposed distributed routing protocols exploiting the novel
{\em store-carry-forward} communication paradigm, within the context
of Delay Tolerant Networking (DTN).
Some of them have already pointed out that node heterogeneity in terms of spatial locality
or inter-contact times can be very beneficial to improve end-to-end delays. In particular, the history of past
encounters \cite{ease,greco1} or the explicit dissemination of information about the
mobility pattern of the nodes \cite{taxi,evaluating} can significantly improve the message delivery
delay without resorting to flooding-based approaches which are very wasteful
of resources.

However, the feasible performance gains, in terms of both throughput and delay,
that can be achieved by exploiting node spatial heterogeneity, as well as the resulting
scaling laws in a network with increasing number of nodes have not been
investigated so far.
% optimality of such routing schemes, as well as the resulting scaling laws
% in a network with increasing number of nodes have not been assessed.
% Indeed, little work has been done in the direction of establishing fundamental
% delay-throughput trade-offs in mobile ad-hoc networks comprising heterogeneous
% nodes.
In this paper, we bridge the theoretical analysis of fundamental
scaling laws of mobile ad-hoc networks with the insights already gained through
practical protocol development. By so doing, we provide a theoretical foundation to the
design of intelligent routing schemes which exploit the spatial heterogeneity of nodes,
analytically showing the potential of such schemes in terms of delay-throughput trade-offs.
%Indeed, the common wisdom matured from throughput-only analysis
%is that node mobility is best exploited under the homogeneous mixing assumption,
%which naturally occurs when nodes move independently of each other
%according to an ergodic process which results into a uniform, stationary distribution over the space,
%such as all theoretical mobility models commonly adopted in the literature.

%In this paper we are going to show that the common belief, according to which
%the best delay-throughput trade-offs are obtained in the case of identical nodes
%uniformly visiting the network area, is, actually, wrong. Node heterogeneity
%and restricted mobility can, indeed, be very beneficial in terms of
%delay-throughput trade-offs, nevertheless this potentiality has been ignored so far.
%also because all mobility models considered so far produce
%a uniform distribution of the node presence over the network.

In particular, we consider a restricted mobility model similar to the one
introduced in \cite{mobihoc07}, in which each node moves around a home-point
randomly chosen in the area.
Nodes move independently of each other, but they are not identical, because each
node is characterized by a different home-point. We consider an ergodic mobility
process which produces, for each node, a spatial stationary distribution
which is rotationally invariant around the home point, and decays
as $d^{-\delta}$, where $d$ is the distance from the home-point, and
$\delta \geq 0$ (the distribution is properly normalized so that its integral
over the area is one). In order that each node achieves the above spatial
distribution we have considered, for simplicity, the reshuffling model, usually referred
to as i.i.d. mobility model.

The family of mobile networks that we consider comprises, as a special case,
the Grossglauser-Tse scenario in which nodes are fully mobile ($\delta = 0$) and therefore
indistinguishable from each other, as well as the limiting case of static nodes
considered by Gupta-Kumar ($\delta \rightarrow \infty$).
We identify a class of scheduling-routing schemes whose performance
in terms of throughput and delay exhibits an intriguing behavior
as we vary $\delta$ (see Figure \ref{fig:deth}).
In particular, for $\delta = 2$ our scheme achieves near-optimal results, i.e.,
almost constant throughput and almost constant delay (except for a poly-logarithmic factor),
and, over a wide range of values for $\delta$, it significantly improves over existing bounds derived
under the assumption of identical nodes.

We emphasize that our schemes do not exploit any sophisticated technique
like the ones that have recently been proposed to improve delay-throughput
trade-offs in the basic cases of Gupta-Kumar and Grossglauser-Tse, such as
hierarchical cooperation with MIMO communications \cite{ozgur} and
source coding \cite{Srikant}, respectively.
The purpose of our work is not to establish optimal information theoretic
results, but to show that there is an additional dimension to be exploited,
i.e., node spatial heterogeneity, which has been so far neglected by theoretical
studies aimed at establishing fundamental scaling laws of mobile ad-hoc networks.
For this reason we will maintain the basic system assumptions
originally introduced by Gupta-Kumar. More sophisticated techniques
can be added to our scheme, and can further improve the bounds
presented here.
\begin{lunga}
The roadmap of the paper is as follows. In Section \ref{sec:notation}
we introduce our system assumptions and notation. In Section \ref{sec:related}
we recall previous delay-throughput results that have been obtained under
assumptions similar to ours. A summary of our new findings,
accompanied by intuitive explanations, is provided in Section \ref{sec:summary}.
In Section \ref{sec:scheme} we describe in details our scheduling-routing schemes
in the fast mobility case, and in Section \ref{sec:analysis} we analyze their performance under the
different regimes that arise while varying the exponent $\delta$.
The slow mobility case is discussed in Section \ref{sec:slowmobility}
We conclude in Section \ref{sec:conclusions}.
\end{lunga}

\section{System assumptions}\label{sec:notation}

\subsection{\bf Mobility model}
We consider a network composed of $n$ nodes moving over a
square region ${\cal O}$ of area $n$ with wrap-around conditions (i.e., a torus),
to avoid border effects.
% a bi-dimensional Torus surface ${\cal O}$ of area $n$.\footnote{A bi-dimensional (topological) Torus
%of area $n$ is the surface generated by the Cartesian product of two circles of
%length $\sqrt{n}$. The Torus can be equivalently described as a quotient of the Cartesian plane
%under the identifications $(x,y) \simeq (x+\sqrt{n},y) \simeq (x,y+\sqrt{n})$.}
% Through this paper we will adopt this second  definition of a Torus.}
Note that, under this assumption, the node density over the area remains constant
as we increase $n$, equal to 1.

Time is divided into slots of equal duration, normalized to 1.
We consider a two-dimensional i.i.d. mobility model,
% introduced in \cite{Modiano-Neely},
according to which the positions of the nodes are totally reshuffled after each slot, independently
from slot to slot and among the nodes.
At the beginning of each slot, a node jumps in zero time to a new position,
and remains in the new position for the entire duration of a slot.

Although the i.i.d. mobility model may appear to be unrealistic,
it has been widely adopted in the literature because of its mathematical
tractability.
\begin{lunga} { \bf We have adopted it in our work especially because it allows to model
in a straightforward way the situation in which the stationary
distribution of a node over the space is not uniform. Indeed, it is
sufficient, at each slot, to sample a random point of the space
according to the desired distribution, and use this point as the new position
of the node during the current slot. Notice that different mobility models,
such as the Brownian motion, random walk and random way-point, do not
permit to obtain in an easy way a desired stationary distribution over
the area.}
\end{lunga}
\begin{corta}
We have adopted it in our work especially because it allows to model
in a straightforward way the situation in which the stationary
distribution of a node over the space is not uniform, which is more
difficult to obtain using other mobility models commonly adopted in the literature.
\end{corta}

Similarly to previous work, we consider two time-scales of node mobility:
\begin{description}
\item{\bf Fast Mobility:} the mobility of nodes takes place at the
same time-scale as packet transmission. Therefore, when two nodes
wish to communicate, they have only one time slot at
their disposal, after which the two nodes separate from each other.
As a consequence, only single-hop transmissions can occur.

\item{\bf Slow  Mobility:} the mobility of nodes is sufficiently
slow that a node can send a packet over multiple hops to reach
another node. This situation is usually modelled by redefining the
time slot as the duration of the \lq coherence interval' during
which the positions of the nodes can be considered to be static.
The topology is reshuffled after each slot\footnote{In the slow mobility case,
the reshuffling (i.i.d.) mobility model can be justified assuming that
devices are disconnected from the network (e.g., switched off) while
travelling from one place to another.}, but several transmissions
can occur during a slot because the packet transmission
time can be set much smaller than the duration of a slot.
% In the extreme case,
% it is possible to perform $\sqrt{n}$ hops in a single slot,
% allowing a node to reach an arbitrarily far destination.
\end{description}

In our work we will consider both fast and slow mobility, but
we warn the reader that the results obtained under these
two scenarios are not directly comparable, because the definition
of time slot is different.

\begin{comment}
In this paper we will consider  a
generalization of two previous models in which $n^\psi$ transmissions can occur in a
time slot with $\psi \in [0,1/2]$. Note that  fast mobility is obtained setting
for  $\psi=0$; while  slow mobility is obtained for  $\psi=1/2$.
\end{comment}

%\item{\bf i.i.D. mobility model}
%\item {\bf Random Way-Point} Each mode  alternates periods in which it keeps still
% (rest phases) to periods in which
%it is moving around. Before starting a new movement phase,  a node selects at
%random a new destination point, a trajectory
% leading to the new destination  and the average speed to be met along its trip;
% then it  moves along the selected trajectory, according
% to an  instantaneous speed profile that meets the selected average  constraint
% (in the simplest cases  trajectories are segments and speed is kept
% constant during the movement phase).
% Once   the new destination has been  reached, the duration of
% the following rest phase is randomly selected by the node.  Duration of rest
% periods dorm a sequence of i.i.d. random variables  distributed according to
% same assigned distribution. We denote with $E[T_M]$ and  $E[T_R]$ respectively
%the average duration of rest periods and moving periods.
%\end{description}

Let $X_i(t)$ denote the position of node $i$ at time $t$ ($t$ is an
integer denoting the slot sequence number) and ${\bf X}(t)=(X_1(t), X_2(t) \ldots X_n(t))$ be
the vector of nodes' positions; we define by $d_{ij}(t)$
the distance between mobile $i$ and mobile $j$ at time $t$, i.e.\footnote{Given any two points
\mbox{$X_1=(x_1,y_1)\!\in\!{\cal O}$} and \mbox{$X_2=(x_2,y_2)\!\in\!{\cal O}$}
we formally define their distance over the torus surface as: $\|X_1(t)-X_2(t)\| = \min_{u,v\in \{-\sqrt{n},0,\sqrt{n}\}}
\sqrt{(x_1+u-x_2)^2+(y_1+v-y_2)^2}$}, \mbox{$d_{ij}(t)=\|X_i(t)-X_j(t)\|$}

Each node $i$ is characterized by a home-point $H_i$, which is uniformly
and independently selected over the area. The collection of the nodes home-points is denoted
by vector ${\bf H} =(H_1, H_2 \ldots H_n)$ and does not change over time,
although it can be different for each network instance.
We define by $d^H_{ij}$ the distance between the
home-points of nodes $i$ and $j$, i.e., \mbox{$d^H_{ij} = \|H_i-H_j\|$}.

The spatial stationary distribution of a node is assumed to be rotationally
invariant with respect to the home-point, and thus can be described by
a generic, non increasing function $\phi(d)$ of the distance $d$ from the
the home-point. In this paper we assume that $\phi(d)$ decays as
a power-law of exponent $\delta$, i.e., $\phi(d)\sim d^{-\delta}$, with $\delta \geq 0$.
This choice is supported by a number of measurements papers which have
found power-laws to be quite ubiquitous in experimental traces related
to both human and vehicular mobility \cite{taxi,levy,castro,evaluating}.
For example, in \cite{castro} authors analyze a large corporate wireless network and
find that the fraction of time spent by users in association to a given access
point exhibits a power law.
\begin{lunga}
For example, in \cite{taxi}
the authors analyze a large mobility trace of taxis
in the city of Warsaw. The empirical distribution of the number of taxis
falling in the cells of a regular grid is found to be heavy-tailed
and fairly stable over time. In \cite{imotes-info06} the duration
of contact times between people is found to be heavy-tailed
in different traces related to conference and campus-wide experiments.
In \cite{castro} authors analyze a corporate wireless local area network and
find that the fraction of time spent by users with a given access
point exhibits a power law.
\end{lunga}

To avoid divergence problems in proximity of the home-point, we take function
$s(d)= \min(1,d^{-\delta})$, and normalize it so as to obtain a proper probability
density function over the network area:
$$\phi(d) = \frac{s(d)}{\iint_{{\cal O}} s(d)}$$
The value of the normalization constant \mbox{$G = {\iint_{{\cal O}} s(d)}$}
can be approximated, in order sense \footnote{Given two functions $f(n)\ge\! 0$ and $g(n)\ge\! 0$: \mbox{$f(n)\!=o(g(n))$} means
$\lim_{n \to  \infty}{f(n)}/{g(n)}=0$; $f(n)=O(g(n))$ means $\limsup_{n \to
\infty}{f(n)}/{g(n)}=c<\infty$; $f(n)=\omega(g(n))$ is equivalent to $g(n)=o(f(n))$;
$f(n)=\Omega(g(n))$ is equivalent to $g(n)=O(f(n))$;
$f(n)=\Os(g(n))$ means  $f(n)=O(g(n))$ and
$g(n)=O(f(n))$; at last  $f(n) \asymp g(n)$ means $\lim_{n\to \infty}
f(n)/g(n)=1$}, by the following integral in polar coordinates:
$$ G = \Theta \left( \int_0^{2 \pi} \diff \theta \int_0^{\sqrt{n}} \min(1,\rho^{-\delta})  \rho \diff \rho \right)$$
We obtain that $G$ is finite for any $\delta > 2$.
For $0 \leq \delta < 2$ we have \mbox{$G = \Theta(n^{\frac{2-\delta}{2}})$}.
For the special value $\delta = 2$ we have $G = \Theta(\log{n})$.
Note that $\delta = 0$ leads to a uniform distribution over the space,
whereas letting $\delta$ go to infinity we obtain the same
behavior as that of a static network.

% Note that, as limit case, we can obtain networks with static nodes; in this case
%$\phi_i(X)=\delta(X-X_i^h)$, being $\delta(X)$ the Dirac impulse
%function. For simplicity, we assume that the mobility of all nodes is
%characterized by the same function $\phi(X)$. However, this restriction can be relaxed
%introducing classes of nodes with different functions $\phi(X)$.
%Indeed our analysis can be easily extended to the case in which
%nodes with different mobility patterns coexist (e.g., a mixture of
%fixed nodes and fully mobile nodes, or several classes of nodes
%with different degrees of mobility around their home-points).

\subsection{\bf Communication model}
To account for interference among simultaneous transmissions,
we adopt the protocol model introduced in ~\cite{Gupta-Kumar}~\footnote{The protocol model
has been proven to be pessimistic with respect to the physical model employing power control
(see Theorem 4.1, pag. 174 in \cite{bibbiakumar}). Thus the results obtained in
this paper can be regarded as lower bounds of the network performance
achievable under the physical model employing power control.}.
% roughly represents the behavior of wireless MAC protocols in the case of
% omni-directional antennas without power capture.
Nodes employ a common range $R$ for all transmissions which occur in the same
time slot; equivalently, they employ a common power level, i.e.,
no power adaptation mechanism is used.
A transmission from node $i$ to node $j$ using transmission range $R$
can be successfully received at node $j$ if and only if the
following two conditions hold:
\begin{enumerate}
    \item the distance between $i$ and $j$ is smaller than or equal to $R$, i.e.,~\mbox{$d_{ij}(t) \leq R$}
    \item for every other node $k$ simultaneously transmitting, \mbox{$d_{kj}(t) \geq (1 + \Delta)\,R$},
    being $\Delta$ a guard factor.
\end{enumerate}
Transmissions occur at fixed rate which is normalized to 1.
We assume that a single copy of each packet is present in the network at any time, i.e.,
data units are not broadcasted nor replicated, and nodes do not keep copies
of previously received packets in their buffer.

\subsection{\bf Traffic model}
Similarly to previous work we consider permutation  traffic patterns in
which $n$ randomly selected source-destination pairs
exchange traffic at rate $\lambda$. Source-destination pairs are
selected is such a way that  every node is origin and destination of a single traffic flow with average rate
$\lambda$. We further assume that, for each pair, the distance between the source home-point
and the destination home-point is $\Theta(\sqrt{n})$,
i.e., we consider the worst case in which all connections are established
among nodes having `far away' home-points.
%\footnote{this can be considered as a worst-case}.
%Without loss of generality, we assume that the destination
%of node $i$ is node $i+1$, and the destination of node $n$ is node $1$.
%We further assume that messages are generated at every source according to
%independent memoryless Bernoullian processes.
%{\bf dove e' che ci serve
%questa ipotesi? non e' un po' restrittiva? non mi pare che sia stata mai fatta
%negli altri paper simili al nostro...}

\subsection{\bf Throughput and delay}
We use the following definitions of asymptotic throughput and delay. Let $L_i(T)$
be the number of packets delivered to the destination of node $i$ in the time interval
$[0,T]$. The delay of a packet is the time it takes for the packet to reach the destination
after it leaves the source. Let $D_i(t)$ be the sum of the delays
experiences by all packets successfully delivered to the destination of node $i$
in the time interval $[0,T]$.
We say that an asymptotic throughput $\lambda$ and an asymptotic
delay $D$ per S-D pair are feasible if there is an $n_0$ such that for any $n \geq n_0$ there exists
a scheduling/routing scheme for which both the following properties hold
\begin{eqnarray*}
\lim_{T \rightarrow \infty} {\rm Pr} \left( \frac{L_i [T]}{T}  \geq \lambda ,  \forall i \right) &=& 1 \\
\lim_{T \rightarrow \infty} {\rm Pr} \left( \frac{D_i [T]}{L_i [T]}  \leq D ,  \forall i \right) &=& 1
\end{eqnarray*}
Equivalently, we say in this case that the network sustains an aggregate throughput
$\Lambda = n \lambda$. We will also adopt the simple Power
Function \cite{power}, defined as the ratio $\lambda/D$,
to characterize the system performance by a single metric.

\section{\!\!\!Previous results for the i.i.d. mobility model} \label{sec:related}
To avoid confusion, we limit ourselves to reporting existing
scaling laws obtained for the i.i.d. mobility model.
We emphasize that all of the following results have been derived
under the assumption that the nodes' spatial distribution
is uniform over the area. %, i.e., $\phi(d)=\frac{1}{n}$.

In \cite{Modiano-Neely} the authors analyze throughput/delay
trade-offs in the case of fast mobility, with or without
packet replication. The following general trade-off
is established:
%\vspace{-2mm}
\begin{equation} \label{eq:dnl}
\lambda=O \left( \frac{D}{n} \right) \vspace{-2mm}
\end{equation}

In particular the two-hop scheme of Grossglauser-Tse is proven to incur a
delay $D= \Theta(n)$ while guaranteeing a per-node throughput
$\lambda= \Theta(1)$.
% study the effect of transmitting redundant packets through multiple paths.
%  For the considered class of schemes it was proved that necessarily the per node
%throughput $\lambda$,  and delivery delay  $D$ experienced by packets are
%are related by:
%\[
%\lambda=O(\frac{D}{n})
%\]
To improve delay, two different schemes exploiting packet redundancy are proposed:
the first is still based on two hops, and achieves $D=\Theta(\sqrt{n})$ and
$\lambda=\Theta(1/\sqrt{n})$; the second employs multiple hops
and achieves better delay performance $D=O(\log n)$ while sacrificing
the per-node throughput $\lambda= O \left(1/(n\log n) \right)$.

The i.i.d. mobility model in the case of slow nodes has been studied
in \cite{Toumpis,Shroff-Lin}. In~\cite{Toumpis} a class of scheduling-routing
schemes that achieves \mbox{$\lambda=\Theta \left( \sqrt{D/n} \log^{-3/2}n \right)$}
was devised. A better trade-off was obtained in \cite{Shroff-Lin}, where the authors show
that necessarily the following relation must hold:
%\vspace{-2mm}
\[
\lambda= O \left(\sqrt[3]{\frac{D}{n}}\log n \right)
\vspace{-2mm}
\]
and they propose a scheme that approaches this bound up to a poly-logarithmic
factor, even in the case of constant delay.
%In particular, the proposed scheme achieves a per node throughput
%$\lambda=\Theta(n^\frac{1}{3}/log n )$ with constant delay was proposed.

Recently it has been shown in ~\cite{Srikant} that previous results can be
further improved by encoding transmitted information at sources.
A class of joint coding-scheduling-routing schemes is introduced
which achieves a trade-off \begin{lunga}:
%\vspace{-2mm}
\begin{equation} \label{eq:srikantfast}
\lambda=\Theta \left(\sqrt{\frac{D}{n}} \right)
\end{equation}
in the case of i.i.d. mobility with fast mobiles, when $D$ is both $\omega(\sqrt[3]{n})$ and
$o(n)$, and
%\vspace{-2mm}
\[
\lambda=\Theta \left( \sqrt[3]{\frac{D}{n}} \right)
\vspace{-2mm}
\]
in the case of i.i.d. mobility with slow mobiles, when $D$ is both $\omega(1)$ and
$o(n)$.
\end{lunga}
\begin{corta}
$\lambda=\Theta (\sqrt{D/n})$ in the case of fast mobiles,
and $\lambda=\Theta (\sqrt[3]{D/n})$ in the case of slow mobiles.
\end{corta}

\section{Summary of results}\label{sec:summary}
A graphical representation of our results is reported in Figure \ref{fig:deth} and
Figure \ref{fig:deth2}. Here we discuss only the fast mobility case. Results for
the slow mobility case are similar and will be presented in Section \ref{sec:slowmobility}.
In Figure \ref{fig:deth} we have plotted, as a function
of $\delta$, the best Power $\lambda/D$ achievable by the class of scheduling-routing
schemes introduced in this paper. We have employed a $log_n$ scale in the vertical axis
so as to show the asymptotic order in $n$. Note that in this scale we can neglect the impact of poly-logarithmic factors.

% Results shown on the graph are indeed achievable by a scheduling/routing
% scheme that will be described later in the paper. Therefore the presented curves can be regarded as
% lower bounds on the system performance.
We observe that the maximum Power according to our scheme is achieved when $\delta = 2$: for this particular
value both $\lambda$ and $D$ remain almost constant with $n$, resulting in a Power which scales as $n^0 = 1$
except for a poly-logarithmic factor. Hence, the performance of our scheme
for $\delta = 2$ is very close (in order sense) to the best result that one can think of.
For $1 < \delta < 2$ the optimal Power has the expression $n^{\frac{-3(2-\delta)}{4-\delta}}$,
whereas for $2 < \delta < 3$ the Power is given by $n^{2-\delta}$.
For all other values of $\delta$ the best Power is equal to $-1$.

% The above two expressions hold in the range of values of $\delta$ for which the resulting Power is larger than
% a minimum which depends on the mobility model: the minimum Power is equal to $-1/2$ in the slow mobility
% case and is equal to $-1$ in the fast mobility case.
% These minimum values are the same as those found in the case of homogeneous nodes: recall
% that the original Gupta-Kumar model has a power of $-1/2$, whereas the original
% Grossglauser-Tse model has a power of $-1$.

For $\delta > 2$ no delay-throughput trade-offs
are possible employing our schemes, since throughput is maximized
jointly with the minimization of delay.
% Indeed a small delay $D = O(\log{n})$ is always achievable
% together with a per-node throughput which scale as $n^{1-\delta/2}/\log{n}$.
% The $\log{n}$ reduction in the per-node throughput can be removed only
% at the expense of a severe degradation of delay.

% the maximum throughput is corresponds also to the minimum delay.
A wide range of delay-throughput trade-offs is instead possible, within our
class of scheduling-routing schemes, for $\delta < 2$. However, for $1 < \delta < 2$,
such trade-offs may result in a sub-optimal Power. This is illustrated in Figure \ref{fig:deth2}, in
which we have plotted (in $\log_n$-$\log_n$ scale) all feasible combinations of $\lambda$ and $D$
for various values of $\delta$. For a given $\delta$, the feasible
combinations of $\lambda$ and $D$ lie on a line
departing from the common point $\lambda = 1$, $D = n$. The little circles, which
lie on the curve $D = \lambda^{-2}$, denote, for each considered $\delta$, the smallest achievable
delay (corresponding also to the smallest throughput), and correspond
to the point in which the Power function is maximum (as reported in Figure \ref{fig:deth}) . Notice that, for
$\delta \leq 1$, all points lie on the curve $D = n \lambda$, having common Power
equal to $-1$. Notice that this result agrees with the general trade-off \equaref{dnl} derived in
\cite{Modiano-Neely} for fast mobiles uniformly distributed in the space.
Restricted mobility starts to be beneficial in terms of delay-throughput trade-offs
when $\delta > 1$. In particular, as $\delta$ approaches $2$, it is possible to push down
the delay towards 1 at the expense of smaller and smaller degradation
of the throughput. For example, for $\delta = 1.95$ one can achieve
a delay close to 1 with just a little penalty in throughput.
We have also shown on the graph the curve $D = n \lambda^2$, % corresponding to \equaref{srikantfast},
which represents the best trade-off available so far for the fast mobility case,
obtained in ~\cite{Srikant}. We observe that our schemes performs better
as soon as $\delta > 4/3$, even without resorting to coding techniques.

\tgifeps{8.5}{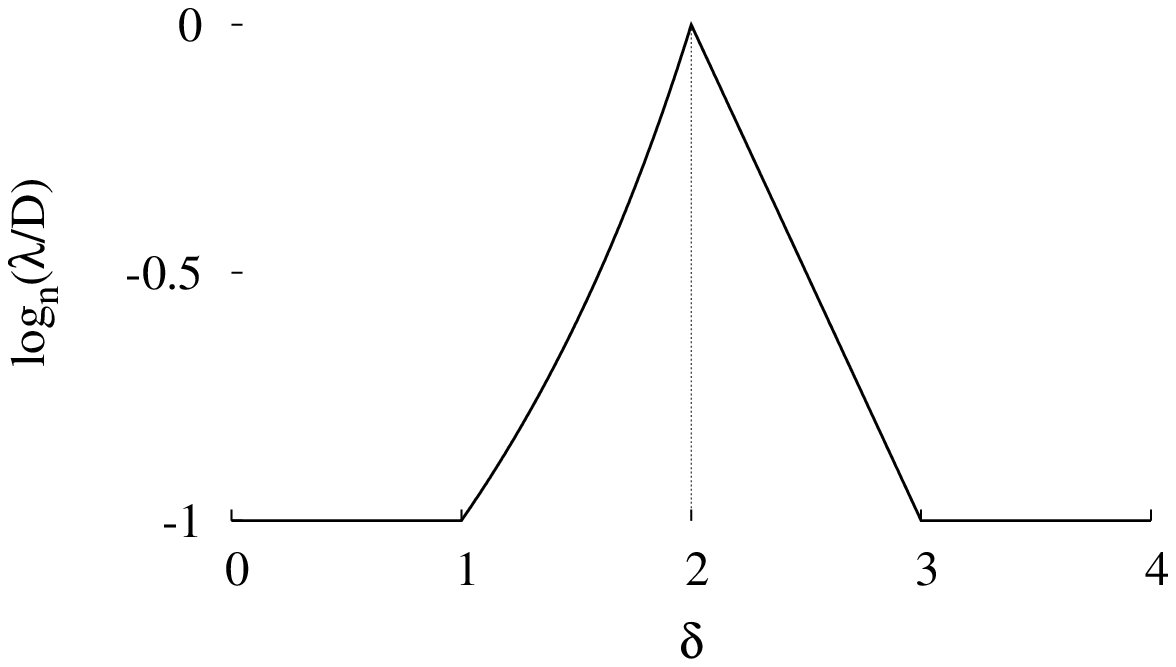}{Power $\lambda/D$ (in log scale) as a function of $\delta$ (fast mobility case).}

\tgifeps{11}{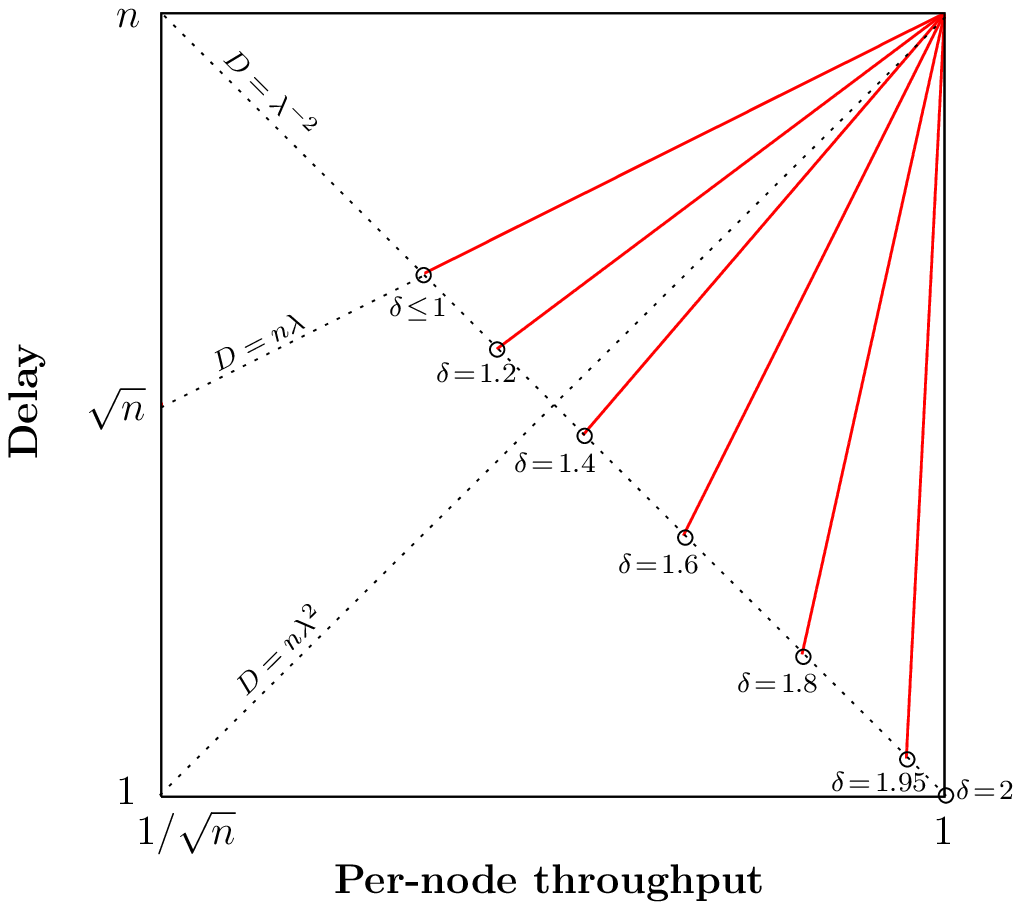}{Delay-Throughput scaling trade-off (in log-log scale)
for different values of $\delta$ between 0 and 2 (fast mobility case).
The marks on the axes represent the orders asymptotically in $n$.}

We now provide an intuitive explanation of the above results.
Our schemes exploit node heterogeneity by an intelligent
selection of relay nodes based on the location of their
home-points\footnote{we assume that each node knows the location of its
home-point and of the home-points of nodes falling in its transmission range.}.
Essentially, we adopt a geographical routing strategy
combined with a divide-and-conquer technique. Data are forwarded
along a chain of relay nodes whose home-points progressively
close in on the home-point of the destination. At each step,
the distance between the home-point of the next-hop relay and
the home-point of the target is halved,
guaranteeing that the message is delivered to the destination in
$\log{n}$ steps. Now, at a given point in space the density of
nodes whose home-points are at distance $O(Z)$ from the point% ~\footnote{Note
% that the number of such nodes is $\Theta(Z^2)$, as shown later in the paper.}
grows as  $Z^{2}Z^{-\delta}/G$. Therefore, for $\delta < 2$ a node
most frequently gets in contact with nodes whose home-point reside far way in
the network area. As a consequence, for $\delta < 2$ the critical step for the system
performance is the last one, in which the packet has to advance by
the minimum distance along the chain of home-points.
Instead, for $\delta > 2$ a node typically gets in contact
with nodes whose home-point are close to the home-point of the node.
As a consequence, for $\delta > 2$ the critical step for the system
performance is the first one, in which the packet has to advance by
the maximum distance along the chain of home-points.
The value $\delta = 2$ is the unique exponent at which the home-points of the
nodes encountered by a given node are equally distributed
{\em at all distance scales}. As a consequence, no step is critical and the system
achieves the optimal performance both in terms of throughput and delay.

We remark that our system presents an interesting analogy with the
problem of navigating small-world graphs using decentralized algorithms employing
only local contact information. In particular we mention here a well known result due to
J.~Kleinberg \cite{kleinberg}, who studied a 2-dimensional lattice enriched
by random shortcuts according to a probability which decays as a power law
with the distance between the connected vertices. The number of hops
required to reach an arbitrary destination exhibits a behavior similar to that
in Figure \ref{fig:deth} as a function of the power-law exponent. In particular, a unique
value of the exponent (equal to 2) allows to navigate the graph in $\log {n}$ hops.

\section{Fast mobility: scheduling-routing schemes}\label{sec:scheme}
In this section we describe the scheduling-routing schemes
that we have devised for our system, in the case of fast mobility.
They form a family of schemes because one parameter, $Z_0$, is free,
and can be specified within a particular scheme so as to achieve a desired delay-throughput trade-off.

Before going on we premise a useful concentration result widely used
in previous work.
\begin{lemma}
Let ${\cal T}$ be a tessellation of ${\cal O}$, whose elements
$T_i$ have area $|T_i|$. The number of home-points falling within each
${T}_i$ is, with high probability, $\Theta( |T_i|)$, uniformly over ${\cal O}$,
as long as $|T_i|=\Omega(\log n)$, $\forall i$.
\label{concentration}
\end{lemma}
We do not repeat the proof of this lemma, which is based on a
standard application of the Chernoff bound (see \cite{motwani}).

\subsection{Routing schemes}\label{subsec:routing}

As already mentioned, we propose a bisection technique that makes
messages advance along a chain of relay nodes whose home points become
progressively  closer to the home point of the destination.
We stop the bisection when the home-point of the target node falls
within distance $Z_0$ from the home-point of the last relay node.

%In presence of  heterogeneity the frequency of node-to node contacts is not
%homogeneous; pairs of nodes whose home-points are
%closer,  come in contact more frequently than pairs with   farther home-points.
%Thus,   to reduce  the expected network delivery delay of data,
%it is convenient for a node $a$, storing a message directed to the final
% destination $d$,  and occasionally in contact with a node $b$,
% to forward its message to $b$,  whenever  $d^H_{ad}> d^H_{bd}$.

%Of course each transmission has a cost in terms of bandwidth;
%thus,  to achieve a good throughput-delay trade off,  on the one
%hand. it  is necessary to convey the information toward the final destination
%through  chains of relaying  nodes (carriers) whose home points become closer
%and closer to the final destination;  on the other hand to limit
%the number of transmissions necessary to reach the final destination.

%For the above reasons we propose a bisection forwarding-scheme according to
%which  all data are guaranteed to be delivered to the final destination
%employing chains of  carriers whose  length is $O(\log n)$.

%{\bf EL - Lo riscritto per renderlo compatibile con ci\`o che e` detto dopo}
%The forwarding scheme works as follows: node $a$  handling   message $m$
%directed to $d$,  first computes $d^H_{ad}=||H_a- H_d||$;

To formalize this idea, suppose that node $a$ handles a
message $m$ directed to $d$. Node $a$ first computes \mbox{$d^H_{ad}=||H_a- H_d||$};
if $d^H_{ad} \le Z_0$ % , where $Z_0$ is a fixed parameter of the scheme,
node $a$ directly forwards the message to node $d$ at the first transmission
opportunity among the two nodes (as dictated by the scheduling scheme).

If $d^H_{ad} > Z_0$, node $a$ compares $d^H_{ad}$ with the set of thresholds
$\{Z_i=2^i Z_0,\; i\in \enne \}$, finding the $i^* \geq 1$ such that,
\mbox{$2^{i^*-1}Z_0 < d^H_{ad} < 2^{i^*} Z_0$}. In this case node $a$ is allowed to
forward the message to any node $b$ whose home-point satisfies the
following geometric constraint with respect to the home-point of $d$:
\mbox{$2^{i^*-2} Z_0 < d^H_{bd} < \gamma~2^{i^*-1} Z_0$}, where $1/2 < \gamma < 1$
is a constant introduced to simplify the analysis of our schemes.
For the sake of concreteness, in the following we will consider the special case
of $\gamma = 3/4$. However, notice that asymptotic results are, in order sense,
insensitive to $\gamma$.
While being handled by $a$, we say that message $m$ is in {\em step} $i^*$ of the
routing algorithm. Note that {\em steps} are counted backward starting from the
destination, and that message $m$ has to go through $i^*$
relay nodes before being delivered to the destination.

\tgifeps{7.5}{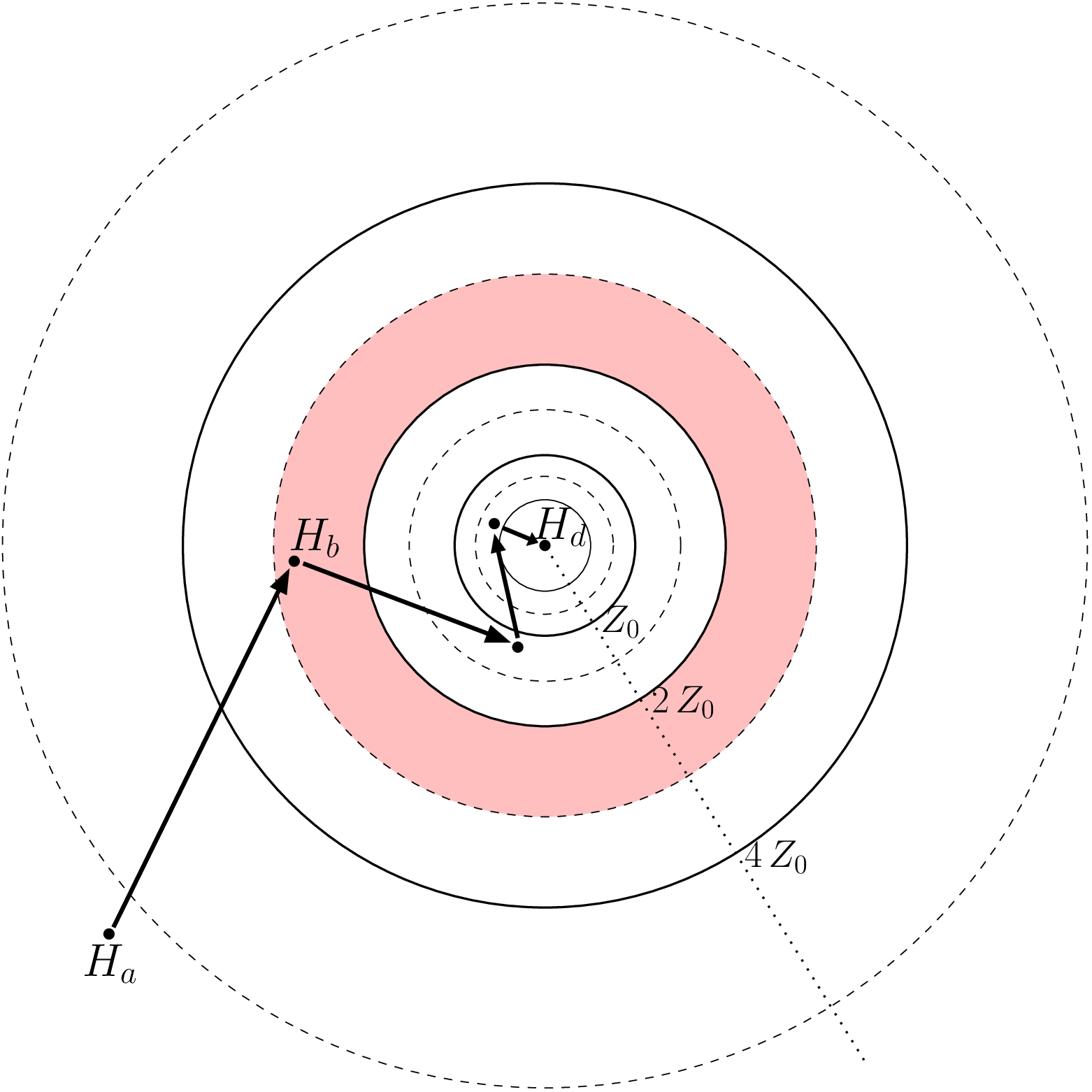}{Example of routing of a message from $a$ to $d$. Dots denote
home-points of the nodes.}

As an example, consider the case illustrated in Figure \ref{fig:routing}, in which source node $a$
wants to send a message to node $d$. Distance $d^H_{ad}$ between the home-points
of $a$ and $d$ satisfies $4 Z_0 < d^H_{ad} < 8 Z_0$, hence we have $i^* = 3$.
This means that message $m$ is in step 3 of the routing algorithm, thus
it has to go through 3 relay nodes before being delivered to the destination.
Since in this case $2^{i^*-2} Z_0 = 2 Z_0$ and $\frac{3}{4} 2^{i^*-1} Z_0 = 3 Z_0$,
the home-points of the nodes $b$ to which $a$ can forward the message
have to lie in the ring $2 Z_0 < d^H_{bd} < 3 Z_0$, which is depicted in Figure \ref{fig:routing}
as a shaded region.

\subsection{Scheduling schemes}\label{subsec:fastscheduling}
A scheduling scheme is in charge of selecting, at each slot, the set of
(non-interfering) transmitter-receiver pairs to be enabled in the network,
as well as the message to be transmitted over each enabled pair.
In our family of schemes, the transmission range employed by a transmitter
depends on the routing step reached by the message to be sent.
To better pack simultaneous transmissions, and thus maximize the
network throughput, our scheduling policy selects, in a given slot,
transmissions having homogeneous transmission ranges. This is done by
selecting, at the beginning of each slot, a step $i$
according to an assigned probability distribution $p^s_i$.
Then the slot is reserved only to messages which are in {\em step} $i$ of the
routing algorithm, i.e., to messages currently stored at nodes whose
home-points are at distances ranging between $2^{i-1}Z_0$ and  $2^{i}Z_0$ from
the home-point of the destination (for the particular case $i = 0$, distances are
between $0$ and $Z_0$).

\tgifeps{6.5}{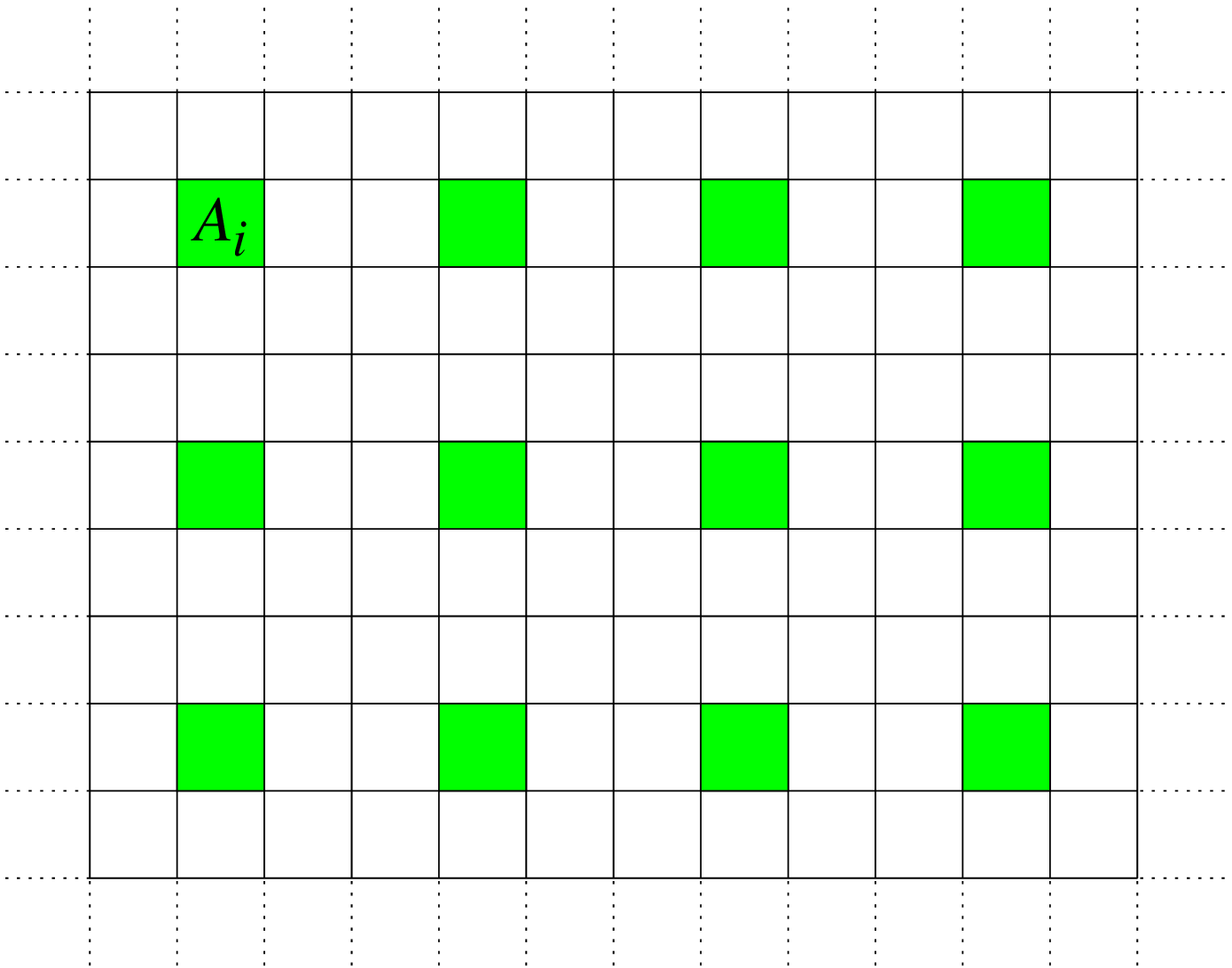}{Example of non-interfering subset of squarelets with $M=9$.}

A common transmission range $R_i$ is employed by all communications
occurring during a slot devoted to step $i$. Once step $i$ has been
selected, the domain ${\cal O}$ is divided into squarelets of area $A_i=R_i^2$,
forming a regular square tessellation. According to the protocol model,
at most one transmission can be enabled in each squarelet.
Moreover, one can easily construct $M = \Theta(1)$ subsets of regularly
spaced, non-interfering squarelets (for example, $M = 9$ assuming a protocol
model with \mbox{$\Delta = 0$}, see Figure \ref{fig:Mnove}).
Each subset can then be enabled in one out
of $M$ slots, guaranteeing fairness among all squarelets and
absence of interference among concurrent transmissions.

\begin{figure*}[bt]
     \centering
     \subfigure[Examples of locations of $H_a$ at routing step 3, given $H_d$ \vspace{-2mm}]{
          \label{subfig:geom1}
                % \psfrag{ylabel}{$1/x$}
                % \psfrag{xlabel}{\small{$x$}}
          \includegraphics[height=1.7in,width=1.7in,angle=0]{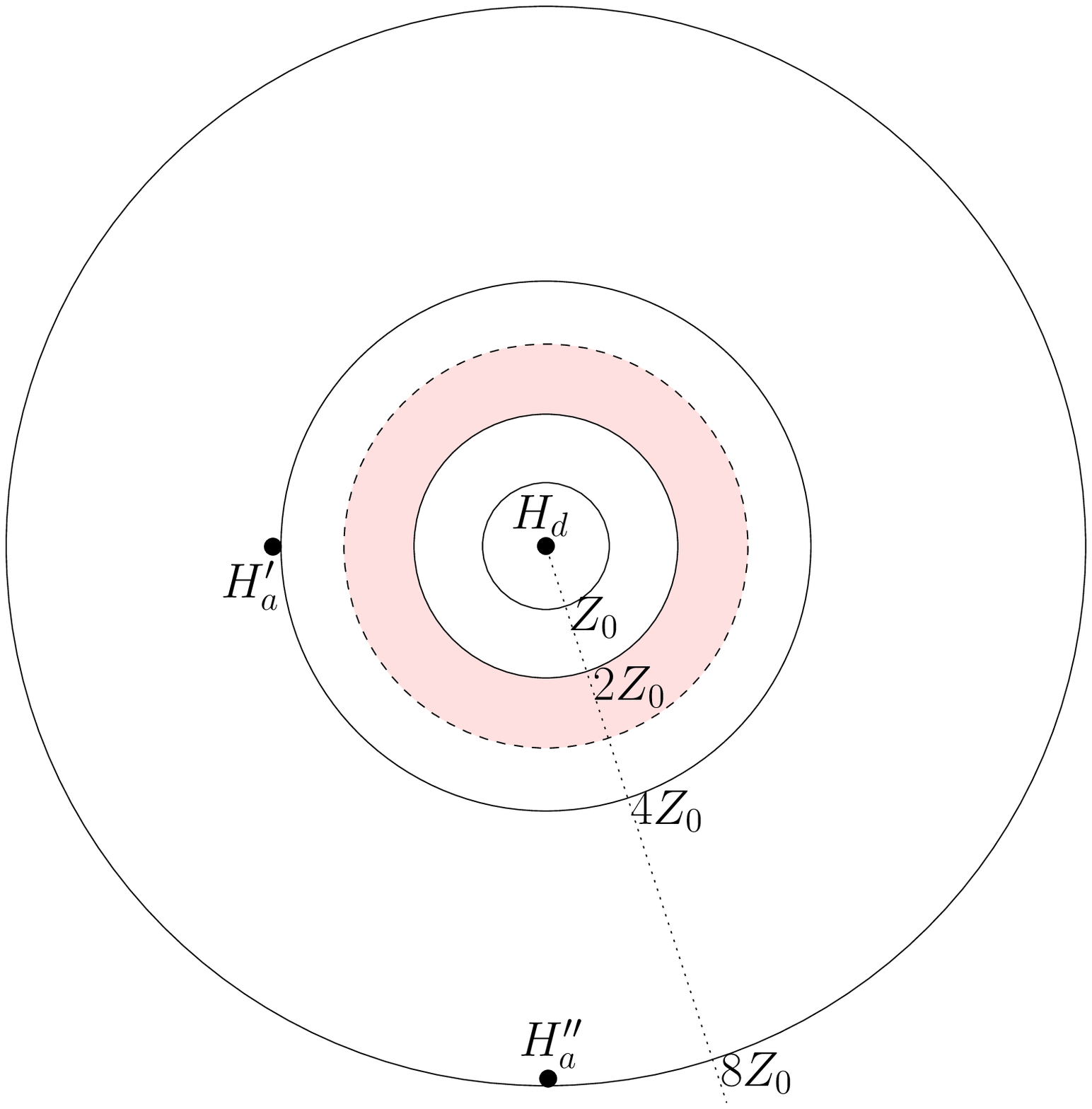}}
     \hfill{}
     \subfigure[Examples of locations of $H_d$ at routing step 3, given $H_a$. The shaded regions
                      around $H'_d$ and $H''_d$ denote the corresponding locations of $H_b$]{
          \label{subfig:geom2}
                % \psfrag{ylabel}{$1/x$}
                % \psfrag{xlabel}{\small{$x$}}
          \includegraphics[height=2.35in,width=2.35in,angle=0]{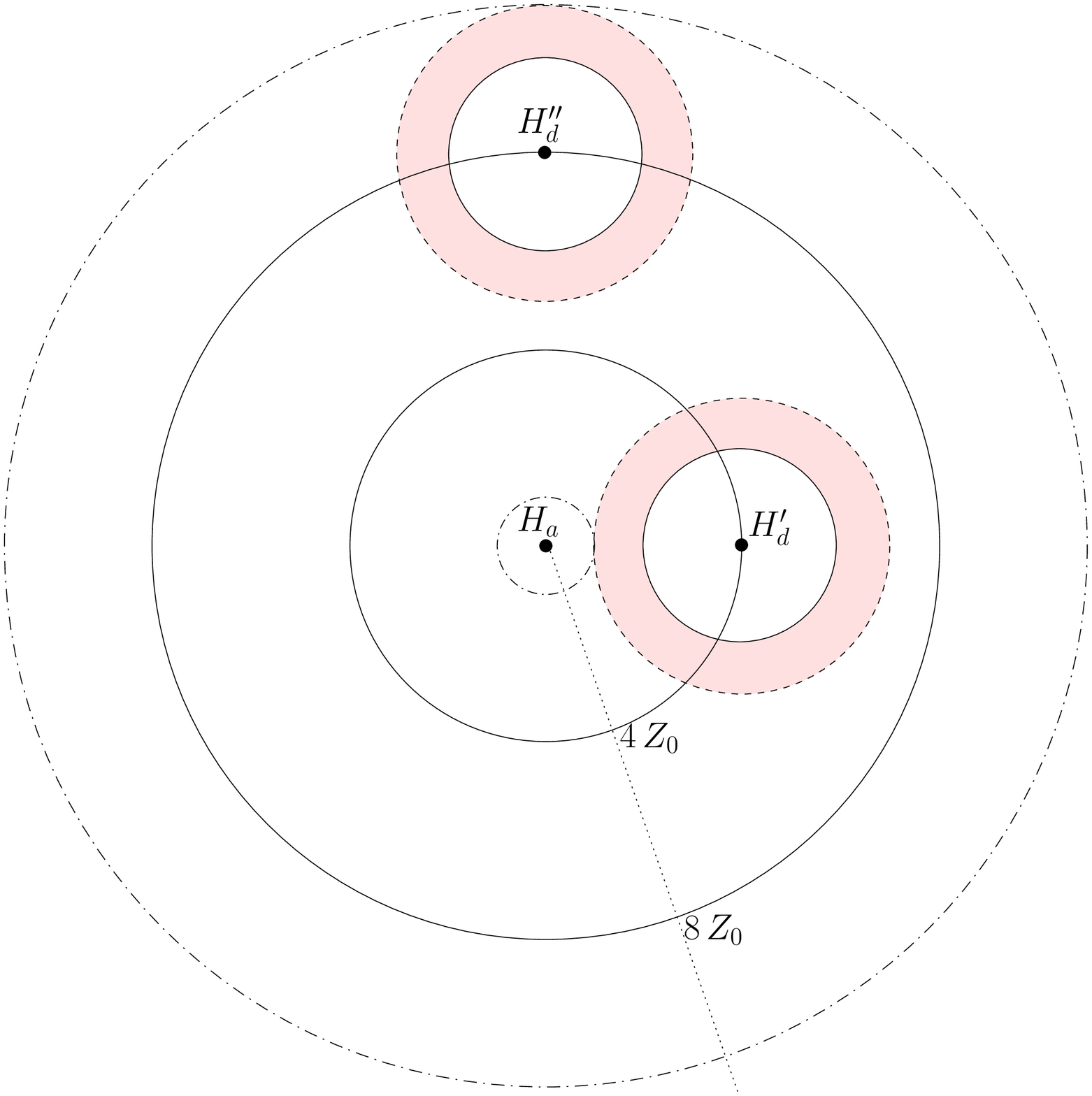}}
     \hfill{}
     \subfigure[The shaded region denotes the union of all possible locations of $H_b$ at routing step 3\vspace{-2mm}]{
          \label{subfig:geom3}
              %  \psfrag{ylabel}{$1/x$}
              %  \psfrag{xlabel}{\small{$x$}}
          \includegraphics[height=2.35in,width=2.35in,angle=0]{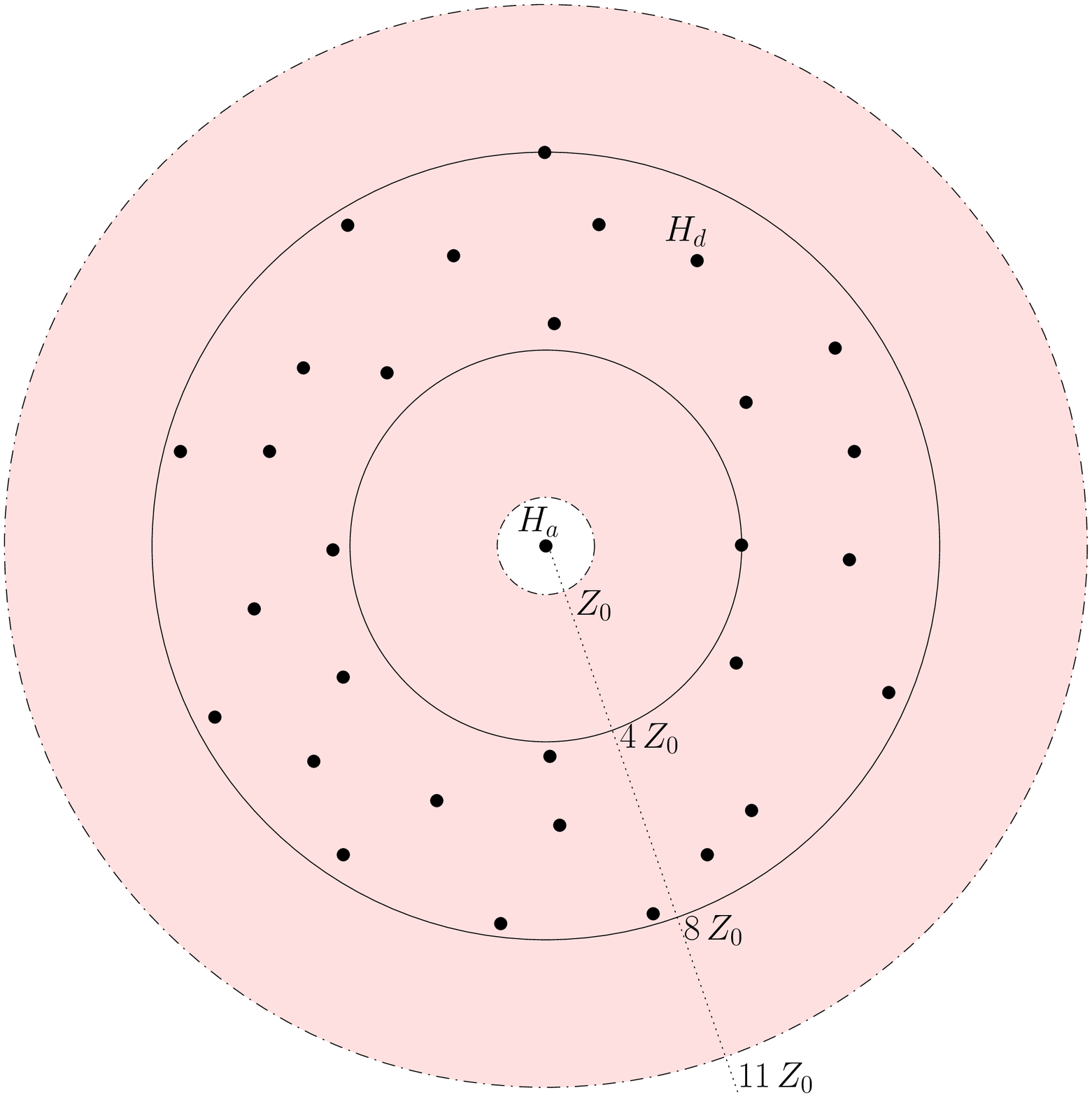}}
     \caption{Geometric considerations supporting the proposed scheduling schemes  \vspace{-2mm}}
%    \vspace{-0.1cm}
    \label{fig:geom}
\end{figure*}

Within a given squarelet ${\cal S}$, the specific transmitter-receiver
pair $(a,b)$ to be enabled is selected as follows.

In slots devoted to the last step ($i = 0$), first all pairs $(a,d)$
residing in ${\cal S}$ and such that: i) $d^H_{ad}<Z_0$;
ii) $a$ has some message $m$ destined to $d$, are first classified
as eligible for transmission. Then, if the set of eligible pairs is not
empty, the squarelet ${\cal S}$ is declared as active and one pair is
randomly selected for transmission.
Notice that, provided that $Z_0=\Omega(\sqrt{\log n})$, by Lemma
\ref{concentration}, each node has w.h.p. $\Theta(Z_0^2)$ destinations $d$.

In slots devoted to step $i>0$, first each node $a$ residing in ${\cal S}$ selects a
message $m$ from its buffer, which has reached step $i$ (if there is any);
let $d_m$ be the final destination of message $m$.
All pairs of nodes $(a,b)$ residing in ${\cal S}$ and such that:
i) $a$ has selected one message $m$; ii) $b$ can act as \mbox{($i\!-\!1$)-th} relay
for message $m$, are first classified as eligible
for transmission.
% i) $a$ can act as $i+1$-th last carrier for message $m_a$;
Then, if the set of eligible pairs is not
empty, the squarelet ${\cal S}$ is declared as active and one pair is
randomly selected for transmission.
% Note that, provided that $Z_i=\Omega(\sqrt{\log n})$, by Lemma \ref{concentration}
% for any message $m$ in step $i$, ($i\ge 1$),
% there are w.h.p. $\Theta(Z_i^2)$ nodes $b$ which may potentially act
% as relay node for messages in step $i$, ($i\ge 1$).
% (all nodes $b$ whose home-points satisfy
% $Z_i/8< d^H_{bd_m}< 11/8 Z_i$).
% and  $d^H_{ab}< 2^{i-1}Z_0$).
We further assume that message $m$ is selected by node $a$ according to a
FIFO scheduling policy.

Thus, logically each node is equipped with a
FIFO queue for each step $i$, $i\ge 1$, plus one queue per destination
for step 0, in which messages are sent directly to the target node (see
Figure \ref{fig:queues}).

An illustration of how our scheduling scheme works is reported in Fig. \ref{fig:geom}.
Suppose, as an example, that the current slot has been assigned to step $i^* = 3$.
Then we have a situation similar to the one already considered
in Figure \ref{fig:routing} while describing the routing scheme.
Figure \ref{subfig:geom1} shows, for a given destination home-point $H_d$,
two possible locations $H'_a$ and $H''_a$ for the home-point of the
transmitting node $a$. In particular, $H'_a$ ($H''_a$) is characterized by
the smallest (largest) distance from $H_d$. If we now take
the point of view of a given transmitting node $a$, we obtain the symmetric
situation depicted in Figure \ref{subfig:geom2}, in which $H'_d$ ($H''_d$)
are located at the smallest (largest) distance from $H_a$. The shaded
ring around $H'_d$ ($H''_d$) denotes the possible location of the home-point
$H_b$ of feasible relay nodes for the corresponding destination.
Notice that, by selecting the head-of-the-line message,
we constrain ourselves to a given destination home-point, and thus we can
only use relay nodes whose home-points lie within a ring similar to the ones depicted
in Figure \ref{subfig:geom2}.

In Figure \ref{subfig:geom3} we have shown the union of all of the possible
rings of the type shown in Figure \ref{subfig:geom2}, as we let the location
of $H_d$ vary. The union region comprises all nodes $b$ satisfying
$\frac{Z_i}{8} \le d^H_{ab}\le \frac{11}{8}Z_i $, for the generic step $i \ge 1$.

Notice again that, provided that $Z_i=\Omega(\sqrt{\log n})$, by Lemma \ref{concentration}
there are w.h.p. $\Theta(Z_i^2)$ nodes $b$ which may potentially act
as relay node, and all of them are at distance $\Theta(Z_i)$ from $a$.
Moreover, the same is true if we constrain ourselves to a given destination
home-point (the area of any shaded ring as in Figure \ref{subfig:geom2} is $\Theta(Z_i^2)$, and all
of its point are at distance $\Theta(Z_i)$ from $a$).
This observation already suggests that our policy can obtain the same
asymptotic performance (in order sense) as the one achievable by a more aggressive
scheduling policy which selects the message to transmit based on the
availability of next-step relays in the same squarelet of the transmitter.
This intuition will be confirmed later in our analysis.

In the following, we will always assume that $Z_0$ (and thus all $Z_i$'s)
are $\Omega(\sqrt{\log n})$, so that we can apply the concentration result in
Lemma \ref{concentration}.

%(notice
%that again provided that $Z_0=\Omega(\sqrt{\log n})$, by proposition
%\ref{concentration},  each node has w.h.p
%$\Theta(Z_0^2)$ last-step destinations).
%In the following, to be able to apply  Proposition \ref{concentration}  we
%will always assume $Z_0=\Omega(\sqrt{\log n})$.

% for any message $m$ in step $i$, ($i\ge 1$),
% there are w.h.p. $\Theta(Z_i^2)$ nodes $b$ which may potentially act
% as relay node for messages in step $i$, ($i\ge 1$).
% (all nodes $b$ whose home-points satisfy
% $Z_i/8< d^H_{bd_m}< 11/8 Z_i$).
% and  $d^H_{ab}< 2^{i-1}Z_0$).

%However, in principle one could use
%Note,  that, by construction, the union of all the possible
%anulus  regions around  all position locations of $H_d$, result in  the
%larger shadowed region represented in  Figure \ref{subfig:geom3}  and comprising
%all  nodes $b$ satisfying  $\frac{Z_i}{4}  \le d^H_{ab}\le
%\frac{7}{4}Z_i $).
% It is immediate to verify that the number of candidate relay nodes
%in each anulus reagion associated to a specific  destination is w.h.p $\Theta(Z^2_{i^*})$, as  the number of  candidate relay nodes
%lying in the larger
%region shadowed in Fig. \ref{subfig:geom3}.

\section{Analysis of the proposed schemes}

\subsection{Design considerations}
Given that step $i$ has been selected at the beginning of a slot, by construction
the number $N_i$ of parallel transmissions that can occur in the network during the slot
equals the number of active squarelets. On average we have:
\begin{equation} \label{eq:eni}
E[N_i]= \frac{n}{M A_i}\Pr\{\mbox{active squarelet}\mid i \}
\end{equation}
where $n$ is the total network area, $M$ is a finite constant accounting for
interference (see \ref{subsec:fastscheduling}) and $\Pr\{\mbox{active squarelet} \mid i \}$
is the probability that a generic squarelet ${\cal S}$ is active at step $i$.

From the discussion in Section \ref{subsec:fastscheduling}, a squarelet ${\cal S}$ is active
if the following two conditions hold: i) at least one pair $(a,b)$ of nodes
is found within ${\cal S}$, such that $b$ can act at as \mbox{($i\!-\!1$)-th}
relay node for $a$ ($i \ge 1$), i.e.,
$\frac{Z_i}{8} \le d^H_{ab}\le \frac{11}{8}Z_i $ (in the case $i = 0$, $b$ must be a destination node satisfying
$\frac{Z_0}{2}  \le d^H_{ab} \le \frac{3}{4}Z_0$); ii) node $a$
has a message to transmit to node $b$ (recall that, for $i \ge 1$, only
the head-of-the-line message in the queue associated to step $i$
can be transmitted).

We observe that the probability that condition i) above holds, denoted
by $\Pr\{\mbox{populated squarelet}\mid i\}$, depends only on the geometry of nodes
and on the mobility process, not on the traffic (i.e., queues dynamics). In general we have
%Note that by construction a necessary condition for a cell  ${\cal S}$ to be
%active in a slot devoted to step  $i\ge 1$, is that at least a pair $(a,b)$ with  $\frac{Z_i}{4}  \le d^H_{ab}\le
%\frac{7}{4}Z_i $ is found within ${\cal S}$
%
% (as shown in Fig.   \ref{fig:geom} ); while in slots devoted to step $i=0$,
% it is necessary to find at   least a pair $(a,b)$ with
% $\frac{Z_0}{}  \le d^H_{ab}\le Z_0 $;
%thus defined with $\Pr\{\mbox{squarelet populated}\mid i\}$  the probability of
%this last event, it results:
\[
\Pr\{\mbox{active squarelet} \mid i\}\le \Pr\{\mbox{populated squarelet}\mid i\}
\]

%We will now describe how to dimension the squarelet sizes $A_i$ in such a way that w.h.p.
%$\Pr\{\mbox{populated squarelet}\mid i\} > 0$.
%Although this criterion provides only an upper bound to the probability that
%a squarelet is active, we will later show in Section \ref{subsec:maximum} that the
%network queues can be loaded, under our schemes,
%in such a way that the probability that node $a$ indeed has a message
%to transmit to $b$ is also w.h.p. greater than zero.

The choice of $A_i$ is critical. The selection of a too large value for $A_i$
leads to a suboptimal exploitation of spatial reuse (thus causing throughput
degradation), without being effective in reducing the delivery delay. This is because
of the contention delay among many eligible transmitter-receiver pairs $(a,b)$ residing in
squarelet ${\cal S}$ (recall that only one of them can be enabled),
which offsets the advantage of reaching a more distant receiver in a single
slot.

On the other hand, the selection of a too small value for $A_i$ is
ineffective in terms of throughput (and also in terms of delay).
This is because the potential benefit of increasing the spatial reuse is countered
by the fact that the fraction of active squarelets tends to 0.

Therefore, a reasonable design choice is to minimize the squarelet size $A_i$
under the constraint that:
\[
 \liminf_{n \to \infty} \Pr\{\mbox{populated squarelet}\mid i\} >0
\]
Although this criterion provides only an upper bound to the probability that
a squarelet is active, we will later show in Section \ref{subsec:maximum} that the
network queues can be loaded, under our schemes,
in such a way that the probability that node $a$ indeed has a message
to transmit to $b$ is also w.h.p. greater than zero.

We will further assume that $\sqrt{A_i}= R_i = O(Z_i)$;
i.e., the transmission range used at step $i$ should be smaller than or equal (in order sense) to the distance
between the home-points of the transmitter and the receiver.
This because it does not make sense to use a transmission range larger (in order sense)
than the distance gained along the chain of home-points
followed by the routing algorithm (otherwise the same transmission range
could be used in a previous step to directly reach the destination,
as done in the last step).

The following lemma will be useful for dimensioning $A_i$.
\begin{lemma}\label{lemma-density}
At routing step $i$, the minimum value of $A_i$ which guarantees
that $\liminf _{n \to \infty} \Pr\{\mbox{populated squarelet}\mid i\} >0$
is given by: %(recall that $Z_i=2^i Z_0$)
\begin{equation} \label{eq:ai}
A_i=\left\{\begin{array}{cc}
\Theta \left(\frac{ n^{1/2}}{Z_i} \right)&   \delta \leq 1\\
\Theta \left( \frac{n^{\frac{2-\delta}{2}}}{Z_i^{2-\delta}}\right) & 1 <\delta<2\\
\Theta \left(\log n      \right)        & \delta=2\\
\Theta \left( Z_i^{{\delta-2}}     \right)           & \delta>2\\
 \end{array}\right.
\end{equation}
\end{lemma}
%First,  observe that  the average number of  potential ($i+1$-th, $i$-th)  carrier pairs
%simultaneously  falling within a squarelet ${\cal S}$,
%$E[N^{\cal S}_{(a,b)}]$,increases quadratically with the surface of ${\cal S}$,
%as result of the fact that   number of potential transmitters $a$ (to which a
%at least receiver corresponds) and
%receivers $b$ (to which a at least transmitter corresponds)
% increases linearly with the surface of ${\cal S}$.
\proof (see Appendix \ref{app:rho}).
\endproof

We observe that $A_i$ depends both on the value
$Z_i=2^i Z_0$ of the associated routing step and on the
power law exponent $\delta$ which characterizes node mobility.

For $\delta < 2$ the largest value of $A_i$
corresponds to the last step (i.e., step 0, having minimum $Z_i$ equal to $Z_0$).
In this case, $Z_0$ cannot be chosen arbitrarily small.
In particular, for any $\delta \leq 1$, condition $\sqrt{A_i}= R_i = O(Z_i)$, coupled
with \equaref{ai}, implies
that $Z_0 = \Omega (n^{1/6})$.
For $1 < \delta < 2$, the same condition implies that
$Z_0 = \Omega \left(n^\frac{2 -\delta}{8-2\delta} \right)$.

For $\delta = 2$, $A_i$ does not depend on $Z_i$, therefore
the same transmission range $R = \sqrt{\log{n}}$ can be used in all steps.
Notice that $R_i = O(Z_i)$ is satisfied because we require
that $Z_0 = \Omega(\sqrt{\log n} )$.

For $\delta > 2$ the first step (having the maximum $Z_i$)
is the one that requires the largest value of $A_i$.
Also in this case, to apply Lemma \ref{concentration}, we require
that $Z_0 = \Omega(\sqrt{\log n} )$.

To summarize, the conditions on the free parameter $Z_0$ of our
class of scheduling-routing schemes are the following:
\begin{equation}\label{eq:condZ0}
Z_0=
 \left\{ \begin{array}{cc}
\Omega \left(n^{1/6} \right) & \delta < 1\\
\Omega \left(n^\frac{2 -\delta}{8-2\delta} \right) & 1 \leq \delta <2\\
\Omega(\sqrt{\log n} ) & \delta \geq 2
\end{array} \right.
\end{equation}

At last, we need to dimension the probability distribution $p^s_i$
with which slots are assigned to the different steps of the routing
algorithm. A natural choice is to equalize of the average
number of transmissions that can occur for each step, and thus avoid that a
particular step becomes the system bottleneck.
This is obtained by making $p^s_i$ inversely proportional to $\overline{N}_i=\frac{n}{A_i}$, which
is an upper bound to the average number of parallel transmissions $E[N_i]$ at step $i$.

Given that $i_{\max}=\lfloor \frac{1}{2}\log_2n-\log_2Z_0 \rfloor$ is the
maximum number of steps traversed by messages, using \equaref{eni} we have
\begin{equation}\label{eq:nobottle}
p^s_i=\frac{\frac{1}{\overline{N}_i}}{\sum_{i=0}^{i_{\max}}\frac{1}{\overline{N}_i}} = \frac{A_i}{\sum_{i=0}^{i_{\max}}A_i}
\end{equation}
We will later see that this choice of $p^s_i$ indeed
equalizes the average number of parallel transmissions occurring at each step.

\subsection{A first throughput characterization}\label{subsec:thro}
An upper-bound to the network throughput achievable by our scheduling-routing schemes
can be easily computed in terms of the maximum number of messages that can flow in one slot
from the sources to the destinations, under the
optimistic hypothesis that $$ \Pr\{\mbox{active squarelet} \mid i\} = \Pr\{\mbox{populated squarelet}\mid i\} $$

% Note that according to our scheduling-routing schemes the network
% load is split evenly (in order sense) among the nodes, thus no congested hot-spots
% arise within the network.

Considering a particular step $i$, the average number of messages stored at
$i$-th relay nodes which can advance to  \mbox{($i\!-\!1$)-th} relay nodes (or be
delivered to the destinations, in the case $i = 0$) in one slot is
bounded by  $p^s_i \overline{N_i}$.
It follows that an upper-bound to the network throughput $\Lambda$ is expressed
by:
\begin{equation} \label{eq:Lambda}
\overline{\Lambda}=n~\overline{\lambda}=\min_i(p^s_i \overline{N}_i)
\end{equation}
Given the design choice \equaref{nobottle} we obtain:
\[
\overline{\Lambda}=p^s_0 \overline{N}_0= \frac{n}{\sum_{i=0}^{i^{\max}} A_i}
\]
For $\delta\le 1$, being  $A_i= \frac{n^{1/2}}{Z_i}$, it follows:
\begin{equation}
\overline{\Lambda}=\frac{n}{\sum_{i=0}^{i_{\max}}\frac{n^{1/2}}{(2^iZ_0)}}
=\Theta(n^{\frac{1}{2}} Z_0)
\label{eq:thrdeltamin1}
\end{equation}
For $1 <\delta < 2$, being  $A_i= \frac{n^{\frac{2-\delta}{2}}}{Z_i^{2-\delta}}$, it follows:
\begin{equation}
\overline{\Lambda}=\frac{n}{\sum_{i=0}^{i_{\max}}\frac{n^{\frac{2-\delta}{2}}}{(2^iZ_0)^{2-\delta}}}
=\Theta(n^{\frac{\delta}{2}} Z_0^{2 -\delta})
\label{eq:thrdeltamin2}
\end{equation}
For $\delta=2$, let $Z_0 = n^\beta$. We have:
\begin{equation}
\overline{\Lambda}=\frac{n}{\sum_{i=0}^{i_{\max}}\log n}=
\left\{  \begin{array}{cc}
\Theta(\frac{n}{\log^{2}(n)}) & \quad \, \beta<1/2\\
\Theta(\frac{n}{\log n}) & \quad  \, \beta = 1/2
\end{array} \right.
\label{eq:thrdelta2}
\end{equation}
At last in the case $\delta>2$
\begin{equation}
\overline{\Lambda}=\frac{n}{\sum_{i=0}^{i_{\max}}(2^iZ_0)^{{\delta-2}}}=\frac{n}{n^{\frac{{\delta-2}}{2}}}
\dfrac{1}{\sum_{i=0}^{i_{\max}}{(2^{{2-\delta}}})^i}=
\Theta(n^{2-\frac{\delta}{2}})
\label{eq:thrdeltamag2}
\end{equation}

\subsection{Delay analysis}\label{subsec:delay}
Now we turn our attention to the delay. We consider a generic node $a$
which is storing a message $m$ in step $i$, directed to final destination $d$.
We need to evaluate the average number of slots needed to make this message
advance by one step. By so doing we neglect the effects of
possible contention for transmission among different messages in step $i$
stored at node $a$; this is equivalent to ignoring queueing delays
within each node, and considering only the service time (we will analyze
queueing effects in Section \ref{subsec:queueing}).

Node $a$ is enabled to transmit message $m$ in a given slot if the
following three conditions hold: i) the slot is assigned to the $i$-th step;
ii) an eligible relay node $b$ for the head-of-the-line message of the
queue devoted to step $i \ge 1$, or a suitable destination node $d$ (for step $i=0$),
is found in the same squarelet in which $a$ resides; iii) among all the eligible
contending pairs residing in the same squarelet, a pair is selected in which $a$
acts as transmitter.
The occurrence of condition i) is simply $p^s_i$.
The probability $p^{\alpha}_i$ of the occurrence of condition ii)
is computed in Appendix \ref{app:pia}:
% Under conditions \equaref{condZ0} on the value of $Z_0$, we have:
\begin{equation}\label{eq:pialfa}
p^{\alpha}_i= \left\{ \begin{array}{cc}
\Theta \left(A_i\frac{|\Gamma_i|}{n} \right) &  \delta \leq 1 \vspace{1mm}\\
\Theta \left(A_i \frac{Z_i^{2(1-\delta)}|\Gamma_i|}{n^{2-\delta}} \right) &  1<\delta<2 \vspace{1mm}\\
\Theta \left(A_i \frac{|\Gamma_i|}{Z_i^{2}} \frac{\log Z_i}{\log^2 n } \right) &  \delta=2 \\
\Theta \left(A_i |\Gamma_i| Z_i^{-\delta} \right) &  \delta>2\\
\end{array}\right.
\vspace{-1mm}
\end{equation}
In the above expressions, $|\Gamma_i|$ denotes the number of potential receivers
of message $m$ that exist in the network. For $i > 0$, we have $|\Gamma_i| =\Theta(Z^2_i)$,
whereas in the last step there exists a unique receiver (i.e., the destination),
hence $|\Gamma_0| = 1$.

The probability $p^{\beta}_i$ of the occurrence of condition iii) depends
on the value of $\delta$. For $\delta < 2$, $p^{\beta}_i = \Theta(1)$,
because the average number $E[N^{\cal S}_{(ab)}]$
of eligible contending pairs $(a,b)$ that can be found in a squarelet is finite (see
Appendix \ref{app:rho}) and by Jensen inequality we
have $p^{\beta}_i> E[1/N^{\cal S}_{(ab)}] > 1/E[N^{\cal S}_{(ab)}] = \Theta(1)$.
For $\delta\ge 2$, instead, there is an infinite number $N^{\cal S}_{a}$ of
transmitting nodes $a$ and a finite number $N^{\cal S}_{b}$ of receiving nodes $b$ (see
Appendix \ref{app:rho}). Therefore, under the pessimistic assumption that all
nodes $a$ have a packet to transmit, it follows $p^{\beta}_i> E [1/N^{\cal S}_{a}]>1/E[N^{\cal S}_{a}] $.
We have, \vspace{-2mm}
\begin{equation}\label{eq:pibeta}
p^{\beta}_i= \left\{ \begin{array}{cc}
\Theta(1) &  \delta < 2\\
\Omega \left(\frac{\log n}{A_i \log A_i} \right) & \delta = 2\\
\Omega(1/A_i) & \delta>2
\end{array}\right.
\vspace{-1mm}
\end{equation}
The probability $p^T_i$ that $a$ transmits message $m$ in a given slot
can then be computed as $p^T_i = p^s_i p^\alpha_i p^\beta_i$:
\begin{equation}\label{eq:ptnotlast}
p^T_i= \left\{ \begin{array}{cc}
\Theta \left(\frac{|\Gamma_i|Z_0}{n^{\frac{1}{2}}Z_i^2} \right) &  \delta \leq 1\\
\Theta \left(\frac{|\Gamma_i|Z_0^{2-\delta}}{Z_i^2n^{1- \frac{\delta}{2}}} \right) &  1 < \delta<2\\
\Omega \left(\frac{|\Gamma_i|\log Z_i}{Z_i^2 \log^3(n)\log\log n } \right) &  \delta=2\\
\Omega \left(\frac{|\Gamma_i|} {Z_i^{2} n^{\frac{\delta}{2}-1}} \right) &  \delta>2\\
\end{array}\right.
\vspace{-1mm}
\end{equation}

At last, under the pessimistic assumption that all nodes in the network are constantly
backlogged by messages in step $i$, the chances that message $m$ is forwarded
in a slot form a memoryless Bernoulli process, since $p^{T}_i$ depends only on the
geometry of nodes, which completely regenerates at every slot.
Thus, it immediately follows that the average service time $D_S^i$ of a message
in step $i$ is equal to $1/p^{T}_i$.
Neglecting queueing delays, the total delay $D_S$ from source to destination
can be computed as $D_S=\sum_i D_S^i$. It follows:
\begin{equation} \label{eq:delay}
D_S=\left\{ \begin{array}{cc}
\Theta( Z_0n^{\frac{1}{2}})  &  \delta \le 1 \\
\Theta( Z_0^{\delta}n^{1-\frac{\delta}{2}})  & 1 \le \delta<2\\
O(\frac{Z_0^{2} \log^3 (n) \log \log n}{ \log Z_0 }) & \delta =2\\
O(n^{\delta/2-1}Z^2_0) & \delta >2
\end{array} \right.
\vspace{-1mm}
\end{equation}

\subsection{The effect of traffic}\label{subsec:queueing}
In the above computation we have assumed that the delay experienced by a message
at a node is of the same order of magnitude as the service time, i.e., the time that
elapses from when the message becomes head-of-the-line to when it is
successfully transmitted.
%The effect of other messages stored within the same node and contending for
%transmission can lead to a delay increase the queueing delay experienced by
%messages. In such conditions the above evaluated delay $D_S^i$ can be
%regarded as the average service time experienced by messages enqueued
%at the $i$-th relay, being $D_S^i$, by construction,
%the time elapsing since message become head-of-the-line to when it is
%successfully transmitted.
This can be justified by Kingman's upper bound to the total delay
of a discrete-time GI/GI/1-FIFO queue \cite{Wolff}, which states that as long as
the second moment $\sigma^2_a$ of the number of simultaneous arrivals is finite,
the second moment $\sigma^2_D$ of service time is also finite and the queue-load  $\rho$ is strictly
less than one, the average queueing delay $D^i$ is bounded by:
\[
D^i\le D^i_S \max \left( 1, \frac{ \sigma^2_a + \sigma^2_{D}}{2 (D^i_S)^2 (1-\rho)} \right)
\]
from which it follows that $D^i = \Theta(D^i_S)$.
In our case the fact that $\sigma^2_a<\infty$ is immediate in light of the
fact that at most one message can be enqueued per slot at any node, while  \mbox{$\sigma^2_{D}<\infty$}
derives from the fact that the service time distribution is stochastically
dominated by a geometric distribution.

\subsection{Maximum Throughput evaluation}\label{subsec:maximum}

\begin{figure}[htb]
\centering
\includegraphics[width=8cm]{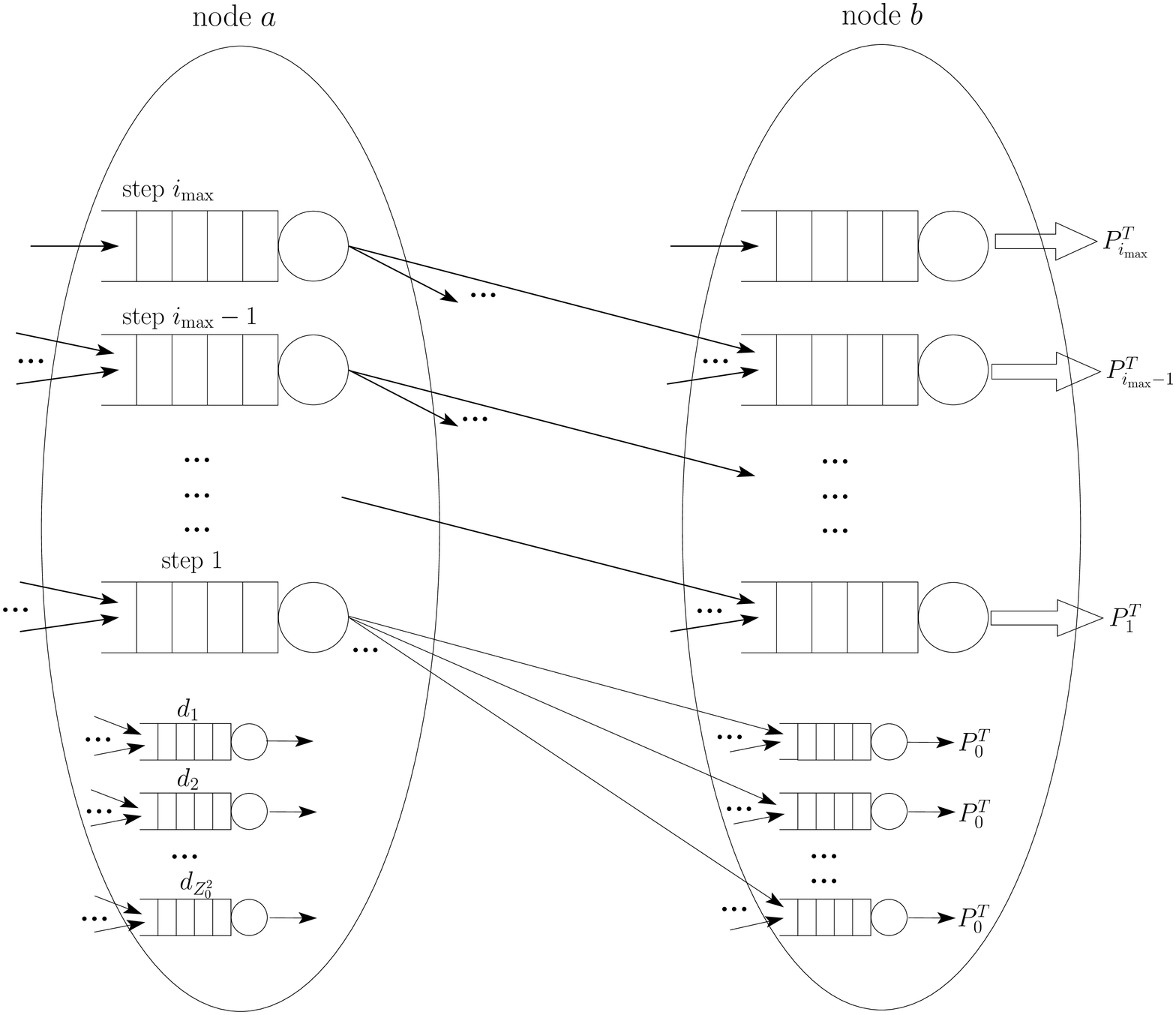}
% \vspace{-1mm}
\caption{Internal architecture of the nodes: queue structure\label{fig:queues}}
% \vspace{-1mm}
\end{figure}

We observe that the whole system can be represented by an
acyclic network of GI/GI/1-FIFO queues; indeed messages advance in the network visiting
queues associated to decreasing step indices $i$, which guarantees the absence of loops (see Figure \ref{fig:queues}).
As a consequence, offered traffic $\Lambda$ can be successfully transferred
through the network, as long as no queue is overloaded.

% \tgifeps{8}{queues}{Internal architecture of a node: queue structure}

The stability condition for the single FIFO queue present at each node $a$, which stores
messages belonging to step $i \ge 1$ of the routing algorithm, is $\lambda^a_i < p_i^T$, where
$\lambda^a_i$ is the aggregate arrival rate of messages in step $i$ at node $a$, while
$p_i^T$ is the transmission probability computed in Section \ref{subsec:delay}.

Considering the last step (i=0), since in this case messages directed to different destinations
are stored in separate queues, the stability condition for the generic queue storing
messages directed to destination $d$ is $\lambda_0^{ad} < p_0^T$,
being $\lambda_0^{ad}$ the arrival rate at node $a$ of last-step messages directed to $d$.

Substituting the expression for $p_i^T$, we immediately obtain that the maximum traffic
sustainable by node queues satisfies:
\[
\lambda^a_i=\Theta(\overline{\lambda})  \qquad i\ge 1 \vspace{-1mm}
\]
and \vspace{-3mm}
\[
\lambda^{ad}_0=\Theta \left(\frac{\overline{\lambda}}{Z_0^2} \right)
\]
i.e., every node $a$ can sustain, for each step, a traffic $\Theta(\overline{\lambda})$,
expressed in messages per slot (in the case of step $i=0$, recall that every node w.h.p.
delivers last-step messages to $\Theta(Z_0^2)$ different destinations).

Note that our scheduling-routing scheme w.h.p. distributes the network traffic
uniformly (in order sense) among all network nodes.
We conclude that the sustainable network throughput
is w.h.p. $\Theta(\overline{\Lambda})$, i.e., the upper bounds previously computed
in Section \ref{subsec:thro} are asymptotically tight.

We remark that this last finding implies  that:
i) no step becomes the system bottleneck under our choice of slot
assignment probabilities $p^s_i$; ii) if the network sustains a throughput
$\Theta(\Lambda)$, necessarily $\Pr\{\mbox{active squarelet} \mid i\}=\Theta( \Pr\{\mbox{populated squarelet}\mid i\})$.

At last, we can claim the optimality of our scheduling-routing schemes
for $\delta<2$. This in light of the following proposition.
\begin{proposizione}
Consider any scheduling-routing scheme according to which messages are
delivered to destinations $d$ by nodes $a$ whose home-points satisfy
$d^H_{ad}=\Theta(Z)$, employing a transmission range $R$.
Then, necessarily the network throughput is $\Lambda=O(\frac{n}{R^2})$.
In addition, if no message replication is employed, the delay is
$D=\Omega \left(1/{\Pr \{ \mbox{meet} \}} \right)$, where
$\Pr \{ \mbox{meet} \} $ is the probability that any two nodes $(a,d)$,
with $d^H_{ad}=\Theta(Z)$, fall within the transmission range of each other.
\label{bounds}
\end{proposizione}
\proof
The throughput condition is an immediate consequence
of the protocol interference model, that prevents two nodes whose relative distance
is smaller than $\frac{\Delta}{2}R$ from transmitting simultaneously \cite{Gupta-Kumar}.
The delay condition immediately descends from the observation that $a$ can deliver
a message to $d$, only if $d$ falls within the transmission range of $a$.
\endproof
Comparing the bounds of Proposition \ref{bounds} with the performance achievable
by our scheduling-routing schemes (for which $Z=Z_0$ and $R=\sqrt{A_0}$), we
conclude that our schemes, for $\delta<2$, achieve optimal delay-throughput trade-offs
within the class of algorithms that do not employ message replication.

\subsection{Possible delay throughput trade-offs}
Let us first examine the case  $\delta> 2$. According to
\equaref{thrdeltamag2} the system throughput does not depend on $Z_0$,
whereas the delay \equaref{delay} increases with $Z_0$.
Thus, $Z_0$ should be minimized to reduce as much as possible
the delay, while no delay-throughput trade-offs are possible.
Recalling the condition \equaref{condZ0} we set $Z_0=\sqrt{\log n}$ and obtain
\[
\lambda= \Theta(n^{1-\delta/2}) \qquad D = O(n^{\delta/2-1} \log n)
\]
Also for $\delta=2$ the dependence of the throughput on $Z_0$ is
weak, since any choice of $Z_0=O(n^\beta)$, with $\beta< 1/2$, leads to the
same throughput, while a $\log n$ factor is gained selecting $Z_0=\Theta(n^{1/2})$,
however at the expense of a severe increase of the delay.
Hence it appears optimal to select $Z_0=\Theta{\sqrt{\log n}}$,
resulting in:
\[
\lambda= \Theta \left( \frac{1}{\log^{2} (n)} \right) \qquad D =  O(\log^{4} (n))
\]
For $\delta<2$, instead, different trade-offs between throughput (optimized
selecting $Z_0$ large) and delay (minimized selecting $Z_0$ small) are possible.
Let $Z_0=n^\beta$.

For $1 <\delta <2$, being $\frac{2-\delta}{8 -2\delta} \leq \beta \leq 1/2$, we obtain
\[
\lambda= \Theta( n^{\delta/2-1+\beta(2-\delta) })\qquad D= \Theta(n^{1-\delta(1/2-\beta)})
\]
whereas for $\delta \le 1$, being $1/6 \leq \beta \leq 1/2$, we obtain
\[
\lambda=  \Theta(n^{\beta-\frac{1}{2} })\qquad D= \Theta(n^{{\beta + \frac{1}{2}}})
\]

\subsection{Alternative scheme for large $\delta$}
It should be noticed that the performance of our scheduling-routing schemes
in the case of $\delta > 2$ rapidly deteriorates in terms of both throughput and delay
for large values of $\delta$. Therefore, we propose here an alternative scheme,
valid for $\delta>2$, whose performance does not depend on $\delta$.

We observe that, for $\delta>2$, every node spends a constant fraction of time
within a finite distance from its home-point. Hence it is possible
to devise a scheduling-routing scheme which is based on this property and that does
not exploit node mobility at all. The network area ${\cal O}$ is divided into a regular
square tessellation whose elements have area equal to $\log{n}$.
Within each squarelet, only the nodes whose home-points belong to the squarelet itself,
are allowed to transmit or receive, using a fixed transmission range
equal to $\sqrt{\log{n}}$.
Since each node spends a constant fraction of time in the squarelet containing
its home-point, the performance that we can achieve is the same as if nodes
were fixed at their home-points, i.e., it is equivalent in order sense to
the Gupta-Kumar case. We can therefore obtain a per node-throughput
$\lambda= \Theta(1/\sqrt{n \log{n}})$ and delay $D_S=\Theta(\sqrt{n/\log{n}})$~\cite{shah1}.

A comparison with our previous scheduling-routing schemes leads to the conclusion
that the alternative scheme becomes convenient for $\delta>3$ (see Figure \ref{fig:deth}).

\section{The slow mobility case}\label{sec:slowmobility}
In presence of slow mobility it is potentially possible to deliver messages
along multi-hop paths in a single slot. Our goal is to understand whether
this possibility can be exploited within our class of scheduling-routing
schemes to further improve their performance.

%We therefore allow an eligible transmitter-receiver pair of nodes
%at the generic step $i$ of the routing algorithm to communicate
%over a multiple-hop path.
A particular case, which serves as a lower bound on the system performance,
is when there is a single step, characterized by $Z_0 = \sqrt{n}$,
for which the resulting scheme essentially degenerates
to the Gupta-Kumar case, providing per-node throughput
$\lambda= \Omega(1/\sqrt{n \log{n}})$ and delay $D_S=\Theta(1)$ (according
to the new definition of slot).

%For $\delta > 2$, the critical step of our scheduling-routing schemes
%is the first one, when $Z_i = \sqrt{n}$. However here there is no advantage
%in using multi-hop paths, because a single hop is enough to reach a candidate
%receiver.

Potentially better delay-throughput trade-offs could be achieved by employing a
hybrid scheme in which messages are first routed according to the previously described
bisection routing scheme, up to arriving at a critical distance from
the destination (to be specified), and then transferred to the destination
in just one slot over a multi-hop path.
This means that when the last step ($i=0$) is scheduled,
the network area is subdivided into squarelets of area
$B_0 = \omega( A_0)$, and many (disjoint) multi-hop transmissions
are established within each squarelet.
This class of hybrid schemes may be effective  for $\delta <2$., i.e.,
 when performance is indeed limited by the last step. Therefore
in the following analysis we assume  $\delta <2$.

\subsection{Analysis}\label{sec:analysis}
We first analyze the throughput performance in the last step, considering
the effect of multi-hop transmissions. Denoting by $Q(B_0)$ the number
of multi-hop transmissions that can be enabled in a squarelet of size
$B_0$ during a slot, and by $\tilde{N}_0$ the total number of parallel
transmissions occurring in the network at step 0, we have
\begin{equation} \label{eq:ntilde}
 E[\tilde{N}_0]= \frac{n}{B_0}E[Q(B_0)]
\end{equation}
In the previous bisection scheme, the squarelet size $A_0$ was dimensioned
in such a way that a finite number of eligible transmitter-receiver pairs
(single-hop) can be found in a squarelet of area $A_0$. Therefore,
in a squarelet of area $B_0$ we have $C_0 = \Theta(B_0/A_0)$ nodes that can
communicate among them over multi-hop paths.
Although the number of pairs that can potentially be established
increases quadratically with $B_0$, only $C_0$ of them
are not conflicting (i.e., do not have a node in common).
Therefore, we have $E[Q(B_0)] = C_0 = \Theta(B_0/A_0)$. Plugging
this expression in \equaref{ntilde} we immediately see
that $E[\tilde{N}_0]= E[{N}_0] = n/A_0$ for any choice of $B_0$,
hence there is no advantage in terms of spatial reuse
by using multi-hop paths.

Turning our attention to the delay of the last step, we consider
a message $m$ stored at $a$ and destined to $d$.
The probability $p_0^\alpha$ that the pair of nodes $(a,d)$
reside in the same squarelet of area $B_0$ is (see appendix \ref{app:pia}):
\[
p_0^\alpha=\left\{\begin{array}{cc} \Theta(\frac{B_0}{n})  & \delta \le 1 \vspace{1mm} \\
\Theta\Big(B_0\frac{Z_0^{2(1-\delta)}}{n^{2-\delta}}\Big)  & 1\le \delta \le 2 \end{array}
\right.
\]
hence it increases linearly with $B_0$. However in this case the probability $p_0^\beta$
that $(a,d)$ is selected is $\Theta(1/C_0) = \Theta(A_0/B_0)$, because
among all $\Theta(C_0^2)$ pairs that can potentially be enabled
only $\Theta(C_0)$ are not conflicting. As a result, the product
$p_0^\alpha p_0^\beta$ does not depend on $B_0$, therefore no gain in terms of
delay is obtained by a scheme that attempts multi-hop
transmissions at the last step. The only exception to previous arguments is represented by the case in which
there is a single step, and thus the hybrid scheme degenerates to
the Gupta-Kumar case.

A graphical representation of the network power achievable in the slow
mobility case is reported in Figure \ref{fig:deth3}.
\tgifeps{7.5}{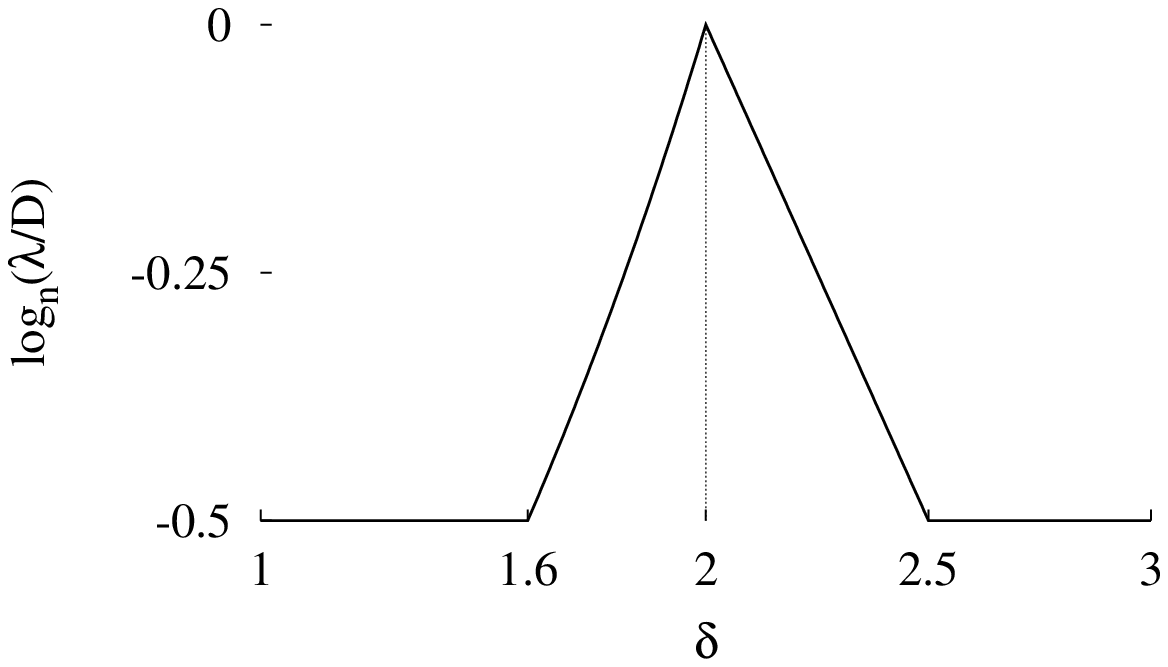}{Power $\lambda/D$ (in log scale) as a function of $\delta$ (slow mobility case).}
We observe that our class of routing-scheduling schemes performs
better than the degenerate case of Gupta-Kumar for values of $\delta$
comprised in the interval $[1.6,2.5]$.

\section{Discussion}\label{sec:discussion}
Although mainly of theoretical interest, we believe that our work
can provide fundamental principles to design smart routing protocols
for mobile ad-hoc networks exploiting the spatial diversity of the nodes.
In particular, our bisection routing strategy appears to be an attractive
solution as long as the spatial distribution of users follows a power law.
Although there is strong experimental evidence that this
is true in many context related to both human and vehicular mobility,
the precise exponent of the power law needs to be carefully estimated,
either through measurements or by an autonomous self-learning
procedure. % (more generally, one has to estimate the mixing degree
%of the nodes in terms of distance scale membership, considering the
%locations of the places most frequently visited by the nodes).
Indeed, two main regimes occur depending on the
exponent value: for values smaller than 2, the larger delays
are expected in the last hop, whereas above 2 the critical hop is the
first one. To avoid large delays, this requires to increase the transmission range
either at the end or at the beginning of the route, which is a rather new concept.
In this paper we have not addressed a number of issues which are essential to
translate our scheme into an implementable solution. In particular, the impact of
more realistic mobility models, and the local dissemination of information
about the home-points of the nodes. However, we believe that the above issues
do not compromise the applicability of our results to a real setting.
Finally, the idea of progressively narrowing the selection of relay nodes
through a divide-and-conquer technique, based on the similarity
of the mobility pattern of the nodes, could be an interesting design
principle to apply in a wider sense than presented here. In this case,
one has to evaluate the mixing degree of the nodes in terms of distance
scale membership, having defined a suitable distance metric between
the mobility pattern of two nodes.

\section{Conclusions}\label{sec:conclusions}
Previous work on delay-throughput trade-offs
in mobile ad hoc networks has mostly been done under the assumption that nodes
are homogeneous and uniformly visit the network area. In this paper we
have shown that this condition can be largely suboptimal. Restricted
mobility, which is usually found in practice, can be exploited by intelligent
scheduling-routing schemes which make use of the geographical
information about the location most visited by a node.
In particular, we have introduced a new class of scheduling
routing schemes which significantly outperform all previously
proposed schemes for a wide range of restricted mobility patterns.

\bibliographystyle{ieee}
% \nocite{*}
%\bibliographystyle{latex8}
\begin{small}

\end{small}
%\end{small}
\ls{0.83}
\appendix

\section{Dimensioning of $A_i$ }\label{app:rho}

We first focus on the last step ($i = 0$).
%However, our results immediately apply to the generic step $i\ge 1$,
%after substituting $Z_0$ with $Z_i$ and $A_0$ with $A_i$.
We consider all distinct pairs of nodes $(a,b)$ whose home-points satisfy
$Z_0/2\le ||H_a -H_b||<3Z_0/4$.

Given a squarelet ${\cal S}$ of area $A_0$, whose center of
mass is for simplicity placed at the origin, we first evaluate
the average number $E[\hat{N}_{(a,b)}^{\cal S}|\bf H]$ of
\lq far' node pairs $(a,b)$ (i.e., pairs such that both $||H_a||>Z_0/4 $ and
$||H_b||> Z_0/4$) that fall within ${\cal S}$.
We further assume $\sqrt{A_i} \le c Z_0$ for some $c < 1/4$.

Let us denote with  $I(H_a,H_b)$ the function that returns 1 if
$Z_0/2\le ||H_a -H_b||<3Z_0/4$ and both $||H_a||> Z_0/4$ and $||H_b||> Z_0/4$.
By construction, it follows:
 \begin{eqnarray*}
 E[\hat{N}_{(a,b)}^{\cal S}|{\bf H}]= \frac{1}{2} \sum_a \sum_{b \neq a} \int_{X_1\in {\cal S}}
 \int_{X_2\in {\cal S}} \phi(H_a-X_1) \\ \phi(H_b-X_2)I(H_a,H_b)   \diff X_1 \diff X_2
 \end{eqnarray*}

Note that  $E[N_{(a,b)}^{\cal S}|\bf H]$ is a random variable over the space of
all the possible home-point locations ${\bf   H}$.
However, since $Z_0= \Omega(\sqrt{\log n})$,  applying  Proposition
\ref{concentration}
it can be immediately proved that, for $n \to \infty$, with high probability, $E[N_{(a,b)}^{\cal S}|{\bf
H}]= \Theta(E[N_{(a,b)}^{\cal S}])$, being $E[N_{(a,b)}^{\cal S}]=E_{\bf H}[E[N_{(a,b)}^{\cal S}]\mid {\bf H}]$
the number of pairs averaged over all possible instances of ${\bf H}$. Hence, we have:
\begin{multline*}
E[\hat{N}_{(a,b)}^{\cal S}] = \frac{1}{2} \int_{X_1\in {\cal S}} \int_{X_2\in {\cal S}} \int_{H_a\in {\cal O}} \int_{H_b\in {\cal O}} \\
\phi(H_a-X_1)\phi(H_b-X_2)I(H_a,H_b) \diff H_a \diff H_b \asymp \\
\asymp \frac{A_0^2 }{2} \int_{H_a\in {\cal O}} \int_{H_b\in {\cal O}} \hspace{-1mm} \phi(H_a)\phi(H_b)I(H_a,H_b) \diff H_a \diff H_b =  \\
= A_0^2\int_{H_a\in {\cal O}} \int_{H_b\in {\cal O}: ||H_b||>||H_a||} \hspace{-12mm} \phi(H_a) \phi(H_b)I(H_a,H_b) \diff H_a \diff H_b = \\
\!\!\!\!\!\!\!\!=\! A_0^2 \!\!\int_{H_a\in {\cal O}: ||H_a||\le CZ_0} \int_{H_b\in {\cal O}: ||H_b||>||H_a||} \hspace{-21mm} \phi(H_a) \phi(H_b)I(H_a,H_b) \diff H_a \diff H_b + \\
\!+ A_0^2\!\!\int_{H_a\in {\cal O}:||H_a|| > C Z_0} \int_{H_b\in {\cal O}: ||H_b||>||H_a||} \hspace{-22mm} \phi(H_a) \phi(H_b)I(H_a,H_b) \diff H_a \diff H_b
\end{multline*}
where the last two terms, for large $C$, may be interpreted, respectively, as
the contribution of pairs $(a,b)$ whose home-points are at distance $\Theta(Z_0)$
and the contribution of pairs whose home-points are at distance $\omega(Z_0)$.

The first term can be approximated by the integral
%\begin{eqnarray*}
%A_0^2  \int_{H_a\in {\cal O}_n:||H_a||<CZ_0}
% \int_{H_b\in {\cal O}_n: ||H_b||>||H_a||}\\
% \phi(H_a)
% \phi(H_b)I(H_a,H_b)dH_a dH_b =\\
%\end{eqnarray*}
\begin{multline*}
\!\!A_0^2\!\int_{H_a\in {\cal O}:Z_0/4<||H_a||<4Z_0} \!\int_{H_b\in {\cal O}: ||H_b||>||H_a||} \hspace{-20mm} \phi^2(Z_0)I(H_a,H_b) \diff H_a \diff H_b = \\
=\Theta\Big(A_0^2 \phi^2(Z_0) Z_0^4 \Big)
\end{multline*}
since, by triangular inequality, it must hold that: \\
$||H_b||<||H_a-H_b||-||H_a||=(C+1)Z_0$

The second term instead provides a contribution
%\begin{eqnarray*}
%A_0^2
%  \int_{H_a\in {\cal O}_n:!|H_a||<4Z_0}
%\int_{H_b\in {\cal O}_n: ||H_b||>||H_a||}\\ \phi(H_a)
% \phi(H_b)I(H_a,H_b)dH_a dH_b=\\
%\end{eqnarray*}
\begin{multline*}
 \Theta\Big(A_0^2 Z_0^2 \int_{H_a\in {\cal O}:||H_a||> C Z_0} \phi^2(H_a) \diff H_a \Big)
\end{multline*}
Substituting the expression for $\phi(X)$, we obtain that the
number of  \lq far' node pairs is:
\begin{eqnarray*}
E[\hat{N}_{(a,b)}^{\cal S}]=\left \{ \begin{array}{cc}
\Theta( \frac{A^2_0 Z^2_0}{n})  & \delta \le 1\\
\Theta(\frac{A^2_0 Z_0^{2(2-\delta)}}{n^{2-\delta}}) & 1<\delta<2\\
\Theta( \frac{A^2_0 }{\log^2 n}) & \delta=2\\
\Theta(A^2_0Z_0^{2(2-\delta)}) & \delta>2\\
\end{array}
\right.
\end{eqnarray*}
To ensure that a far pair $(a,b)$ is found within ${\cal S}$ with a non-vanishing
probability it is necessary to dimension $A_0$ in such a way that $E[\hat{N}_{(a,b)}^{\cal S}]$ is
not vanishing as $n \to \infty$. On the other hand making $E[\hat {N}_{(a,b)}^{\cal S}]=\Theta(1)$ is
also sufficient to guarantee that a far pair is found within squarelet ${\cal S}$
with non vanishing probability.

Now we turn our attention to \lq near' pairs, i.e., pairs of nodes $(a,b)$
in which one of the two nodes (let it be node $a$) has home-point satisfying:
$||H_a||< Z_0/4$. First, note that by triangular inequality it must be
\mbox{$Z_0/4<||H_b||< Z_0$}.% , whenever \mbox{$||H_a||< Z_0/4$}.
Thus to find a \lq near' pair  $(a,b)$ within ${\cal S}$ with
finite probability, we must find within ${\cal S}$ at least one node $a$
whose home-point satisfies  $||H_a||<Z_0/4$ and at least one node $b$
whose home-point satisfies  $ Z_0/4<||H_b||< Z_0$.

The average number $E[N_a]$ of near nodes $a$ in squarelet ${\cal S}$ can be evaluated as:
\begin{eqnarray*}
\!E[N_a]\!\asymp\! A_0\!\int_{H_a\in {\cal O}:||H_a||< \frac{Z_0}{4}} \hspace{-14mm} \phi(H_a) \diff H_a =
\left\{\begin{array}{cc}
\Theta( \frac{A_0 Z_0^{2-\delta}}{n^{\frac{2-\delta}{2}}})& \delta <2\\
\Theta(A_0\frac{\log Z_0 }{\log n})& \delta =2\\
\Theta(A_0) & \delta >2\\
\end{array}\right.
\end{eqnarray*}
whereas the average number $E[N_b]$ of nodes $b$ within ${\cal S}$,
such that $1/4 Z_0<||H_b||< 5/4Z_0$, is
\begin{eqnarray*}
\!\!\!E[N_b]\!\asymp\!A_0\!\!\int_{H_b\in {\cal O}:\frac{Z_0}{4} < ||H_b||< Z_0} \hspace{-16mm} \phi(H_b) \diff H_b =
\left\{\begin{array}{cc}
\Theta( \frac{A_0 Z_0^{2-\delta}}{n^{\frac{2-\delta}{2}}})& \delta <2 \vspace{1mm}\\
\Theta(\frac{A_0 }{\log n})& \delta =2\\
\Theta(A_0 Z_0^{2-\delta}) & \delta >2\\
\end{array}\right.
\end{eqnarray*}
In order to have near pairs within squarelet ${\cal S}$ with a non vanishing
probability, both $E[N_a]$ and $E[N_b]$ must be non vanishing.
At last, $A_0$ must be dimensioned in such a way that either both $E[N_a]$ and $E[N_b]$
are non-vanishing, or $E[\hat{N}_{(a,b)}^{\cal S}]$ is non-vanishing, leading to the
expressions in \equaref{ai}. for  $i=0$.

Now turning our attention to the case $i>0$, we have to evaluate the average
number of distinct pair of nodes $(a,b)$, whose home-point satisfy
$\frac{1}{8}Z_i \le ||H_a -H_b||\le \frac{11}{8}Z_i$,
falling within  a squarelet ${\cal S}$ of area $A_i$, whose center of
mass is for simplicity placed at the origin.

This can be done using
exactly the same approach as before, i.e., distinguishing between \lq far' node pairs $(a,b)$
 (pairs such that both $||H_a||\ge  Z_i/16 $ and $||H_b||\ge  Z_i/16$) and `near'
node  pairs (pairs $(a,b)$ in which at least one of the two nodes (let it be node $a$) has the
home-point satisfying $||H_a||< Z_i/16 $), and evaluating separately the two
contributions.

\section{Evaluation of $p^{\alpha}_i$}\label{app:pia}

Consider a pair of nodes $a$ and $b$ such that $||H_a-H_b||=D$. We
wish to evaluate (in order sense) the probability  $p_{(a,b)}$ that $(a,b)$ are found
within the same squarelet ${\cal S}$ of area $A$:
\begin{multline*}
\lefteqn{p_{(a,b)}=\sum_{\cal S}\Pr\{ a \in {\cal S}\}\Pr\{ b \in{\cal  S}\} = } \\
\sum_{\cal S} \int_{X_1\in{\cal  S}}\phi(X_1-H_a) \int_{X_2\in{{\cal S}(X_1)}}\phi(X_2-H_b) \diff X_2 \diff X_1 = \\
= \int_{X_1\in {\cal O}}\phi(X_1-H_a) \int_{X_2\in {\cal  S}(X_1)}\phi(X_2-H_b) \diff X_2 \diff X_1
\end{multline*}
being ${\cal S}(X_1)$ the squarelet containing $X_1$.
Let $ {\cal V}^a \subset {\cal O}$ denote the set of points $X$
closer to $H_a$ than $H_b$ (i.e., such that: $||H_a -X||<||H_b-X||$),
then by symmetry:
\begin{multline*}
\lefteqn{p_{(a,b)}= 2\int_{X_1\in {\cal V}^a} \hspace{-5mm} \phi(X_1-H_a)
\int_{X_2\in {\cal  S}(X_1)} \hspace{-10mm} \phi(X_2 - H_b) \diff X_2 \diff X_1= } \\
= 2\int_{X_1\in {\cal V}^a: ||X_1-H_a||<\frac{D}{4} } \hspace{-10mm} \phi(X_1-H_a)
\int_{X_2\in {\cal  S}(X_1)} \hspace{-10mm} \phi(X_2-H_b) \diff X_2 \diff X_1 + \\
+ 2\int_{X_1\in {\cal V}^a:  \frac{D}{4} \le ||X_1-H_a||< 4D} \hspace{-10mm} \phi(X_1-H_a)
\int_{X_2\in {\cal  S}(X_1)} \hspace{-10mm} \phi(X_2-H_b) \diff X_2 \diff X_1 + \\
+ 2\int_{X_1\in {\cal V}^a: ||X_1-H_a|| \ge 4D} \hspace{-10mm} \phi(X_1-H_a)
\int_{X_2\in {\cal  S}(X_1)} \hspace{-10mm} \phi(X_2-H_b) \diff X_2 \diff X_1
\end{multline*}
Now assuming $\sqrt{A} < c~D$, being  $c \ll 1$,
%\begin{eqnarray*}
%2\int_{X_1\in {\cal V}^a: ||X_1-H_a||<\frac{d}{4} } \hspace{-10mm} \phi(X_1-H_a)
%\int_{X_2\in {\cal  S}(X_1)} \hspace{-10mm} \phi(X_2-H_b) \diff X_2 \diff X_1 = \\
%= \Theta \left( A \phi(d) \int_{||X_1||< \frac{d}{4}} \phi(X_1) \diff X_1 \right)
%\end{eqnarray*}
the first term in the above expression provides a contribution
\begin{eqnarray*}
\Theta \Big( A \phi(D) \int_{||X_1||< \frac{D}{4}} \phi(X_1) \diff X_1 \Big)
\end{eqnarray*}
since by triangular inequality on the considered domain, at the same time
$||H_b-X_1||\ge||H_a-H_b||-||H_a-X_1||\ge ||H_a-H_b||-||H_a||-||X_1||> 3D/4-cD$  and
$||H_b-X_1||<||H_a-H_b||+||H_a-X_1||<5D/4+cD$, while $||X_2-X_1|| <c~D$.

The second term provides instead a contribution $\Theta \left( A D^2 \phi^2(D) \right)$
%\begin{eqnarray*}
%2\int_{X_1\in {\cal V}^a_n: d/4\le ||X_1-H_a||<4d
%}\phi(X_1-H_a)\\
%\int_{X_2\in {\cal  S}(X_1)}\phi(X_2-H_b) dX_2 dX_1= \Theta( A d^2 \phi^2(d))
%\end{eqnarray*}
since on the considered domain it must be $D/4<||H_b-X_1||<5D+c$.

At last the third term provides a contribution
% \begin{eqnarray*}
%   2\int_{X_1\in {\cal V}^a_n:   ||X_1-H_a||>4d
%}\phi(X_1-H_a)\\
%\int_{X_2\in {\cal  S}(X_1)}\phi(X_2-H_b) dX_2 dX_1= \\
%\Theta\Big(A \int_{X_1\in {\cal V}^a_n:
%||X_1-H_a||>4d}\\
%\Phi^2 (X_1-H_a) dH_a \Big)
% \end{eqnarray*}
\begin{eqnarray*}
\Theta \Big( A \int_{X_1\in {\cal V}^a: ||X_1-H_a||>4D} \phi^2 (X_1-H_a) \diff H_a \Big)
\end{eqnarray*}
since on the domain $||H_a-X_1||=\Theta( ||H_b-X_2||)$.

% Note that $p_{(a,b)}$  asymptotically depends  only on $d$.
Substituting the expression for $\phi(X)$ we obtain:
\[
p_{(a,b)}=\left \{ \begin{array}{cc}
\Theta(\frac{A}{n})  & \delta< 1 \vspace{1mm} \\
\Theta(A \frac{D^{2(1-\delta})}{n^{2-\delta}})  & 1<\delta< 2 \vspace{1mm}\\
\Theta(A \frac{D^{-2}\log D }{\log^2 n}) & \delta= 2\\
\Theta(A D^{-  \delta})  & \delta>2
\end{array} \right.
\]
At last, observe that  $p_{(a,b)}$ equals $p_0^\alpha$ if we set
\mbox{$Z_0/2<D<3Z_0/4$}, and either $A=A_0$ in case of fast  mobility or
$A=B_0$ in case of  slow  mobility.

Now, to evaluate the expression of $p_i^\alpha$ for $i\ge 1$,
let $\Gamma_i$ denote the set of all potential nodes $b$ which can receive
a message in step $i$ from node $a$. By definition:
\begin{multline*}
\lefteqn{p_i^\alpha = \sum_{\cal S} \Pr\{ a \in {\cal S}\} \Pr\{ \exists b \in \Gamma_i: b \in {\cal S} \} =} \\
= \sum_{\cal S} \Pr\{ a \in {\cal S}\} \Big( 1 - \prod_{b \in \Gamma_i}(1 - \Pr\{ b \in {\cal S}\}) \Big)
 \end{multline*}

In addition, since it can be proved (as shown below)  that:
\begin{equation}
\Big( 1 - \prod_{b \in \Gamma_i}(1 - \Pr\{ b \in {\cal S}\})
\Big)=\Theta\Big(\min(1,\sum_{b\in \Gamma_i}\Pr\{ b \in {\cal S}\})\Big)
\label{asympt-proof}
\end{equation}
we can conclude that $p_i^\alpha =$
  \begin{multline*}
\!\!\!\!\!\!\!\!=\! \Theta \!\left( \sum_{\cal S} \! \int_{X_1\in{\cal  S}} \hspace{-5mm} \phi(X_1-H_a)
\min \! \Big( 1, \sum_{b \in \Gamma_i} \int_{X_2\in{\cal S} } \hspace{-5mm} \phi(X_2-H_b)
\diff X_2 \! \Big) \!\diff X_1 \!\! \right)
 \end{multline*}
Then operating in a way similar to the computation
of $p_{(a,b)}$ we obtain the expressions in \equaref{pialfa}.

To conclude the proof we have to show how to obtain (\ref{asympt-proof}).
 The proof if trivial if there is at least one $b^*\in \Gamma_i$  such
that $\Pr\{ b^* \in {\cal S}\})=\Theta(1)$, since by construction
\[
\Big( 1 - \prod_{b \in \Gamma_i}(1 - \Pr\{ b \in {\cal S}\})
\Big)> \Pr\{ b^* \in {\cal S}\}
\]
So, suppose that $\Pr\{ b \in {\cal S}\})=o(1)$ for all $b \in \Gamma_i$.
In this case, first note that
 $$ \log  \prod_{b \in \Gamma_i}(1 - \Pr\{ b \in {\cal
S}\})= \sum _{b \in \Gamma_i} \log (1- \Pr\{ b \in {\cal S}\})$$
Then, since   \mbox{$- 2x < \log (1-x)\le x$} for any $0 \le  x \le x_0$,
with $x_0\approx 0.8$ solution of equation $1-x_0=\exp(-2x_0)$, it follows:
\[
- 2 \!\sum _{b \in \Gamma_i} \Pr\{ b \in {\cal S}\} \le \log  \prod_{b \in \Gamma_i}(1 - \Pr\{ b \in {\cal S}\})
\le - \!\!\sum _{b \in \Gamma_i} \Pr\{ b \in {\cal S}\}
\]
for $ \sum _{b \in \Gamma_i} \Pr\{ b \in {\cal S}\}\le x_0$, from which:
\begin{multline}
1- \exp(-\sum _{b \in \Gamma_i} \Pr\{ b \in {\cal S}\} )\le 1-  \prod_{b \in
\Gamma_i}(1 - \Pr\{ b \in {\cal S}\}) \le \\\le
1- \exp(- 2\sum_{b \in \Gamma_i} \Pr\{ b \in {\cal S}\} )
\end{multline}
Now $ 1- \exp(-\sum _{b \in \Gamma_i} \Pr\{ b \in {\cal S}\} )\ge 1/2\sum _{b
\in \Gamma_i} \Pr\{ b \in {\cal S}\}$ for $\sum _{b \in \Gamma_i} \Pr\{ b \in
{\cal S}\}< x_0$, thus
\[
 1/2\sum _{b
\in \Gamma_i} \Pr\{ b \in {\cal S}\} \le  1-  \prod_{b \in
\Gamma_i}(1 - \Pr\{ b \in {\cal S}\}) \le 2 \!\!\sum_{b \in \Gamma_i} \Pr\{ b \in {\cal S}\}
\]
i.e., the assertion is proved for $\sum _{b \in \Gamma_i} \Pr\{ b \in {\cal S}\}<
x_0$.
The extension to the case $\sum _{b \in \Gamma_i} \Pr\{ b \in {\cal S}\}\ge
x_0$ is trivial in light of the fact that $( 1 - \prod_{b \in \Gamma_i}(1 - \Pr\{ b \in {\cal S}\}))$
is increasing with respect to $ \Pr\{ b \in {\cal S}\}$.

\begin{comment}
\[
p^\alpha_i=\left \{ \begin{array}{cc}
\Theta(A_i \frac{Z_i^2}{n})  & \delta< 1\\
\Theta(A_i \frac{Z_i^{2(2-\delta})}{n^{2-\delta}}) & 1<\delta< 2\\
\Theta(A_i \frac{\log d}{\log^2 n}) & \delta= 2\\
\Theta(A_i Z_i^{(2-  \delta)})  & \delta>2
\end{array} \right.
\]
\end{comment}

\end{sloppypar}
\end{document}